\documentclass[12pt, a4paper]{article}
\usepackage[height=22cm,width=16cm, centering]{geometry}
\usepackage{setspace}

\usepackage[utf8]{inputenc}

\usepackage{amsmath}
\usepackage{amssymb}
\usepackage{graphicx}
\usepackage[subrefformat=parens]{subcaption}
\usepackage{color,comment, bm, ascmac, microtype, fonttable}
\usepackage[sort&compress, numbers, merge]{natbib}

\definecolor{purple}{rgb}{0.5 ,0, 0.7}
\definecolor{bluegreen}{rgb}{0, 0.5, 0.7}
\usepackage{hyperref}
\hypersetup{colorlinks=true, linkcolor=bluegreen, citecolor=purple, urlcolor=blue}

\DeclareMathOperator{\arcosh}{arcosh}
\DeclareMathOperator{\arsinh}{arsinh}
\DeclareMathOperator{\artanh}{artanh}

\usepackage{acro}
\DeclareAcronym{gr}{
    short=GR ,
    long=general relativity
}
\DeclareAcronym{nec}{
    short=NEC ,
    long=null energy condition
}

\newcommand{\CosmoLattice}{$\mathcal{C}$osmo$\mathcal{L}$attice}
\newcommand{\Thalf}{\mathnormal{\Delta} t_\phi} % half-period of phi oscillation
\newcommand{\Thalfa}{\mathnormal{\Delta} t_a} % half-period of $a$ (scale-factor) oscillation
\newcommand{\Ne}{N_\mathrm{e}} % e-folding number 
\newcommand{\Nc}{N_\text{c}} % Number of color
\newcommand{\alphaCL}{\alpha_{\mathcal{C}\mathcal{L}}} % alpha-time parameter of the CosmoLattice

\newcommand{\colorBigCrunchPositive}{red}
\newcommand{\colorBigCrunchNegative}{purple}
\newcommand{\colorCyclic}{sky blue}
\newcommand{\colorShortInflation}{green}
\newcommand{\colorLongInflationTachyonic}{orange}
\newcommand{\colorLongInflation}{yellow}

\newcommand{\colorChoice}{white}
\newcommand{\colorFlatness}{black}

\allowdisplaybreaks[4]

\begin{document}

\begin{titlepage}

\vspace*{-16 mm}
\begin{flushright}
{\scriptsize 
YITP-23-60 \\
CTPU-PTC-23-15 \\
TU-1187 \\
}
\end{flushright}

\vspace*{.1cm}
\begin{center}
{\Large \bf Dissipative Emergence of Inflation \par \vspace{2mm} from Quasi-Cyclic Universe}\\

\vspace*{.6cm} 
{\large 
Hiroki Matsui,$^{a, }$\footnote{\tt hiroki.matsui@yukawa.kyoto-u.ac.jp} 
Alexandros Papageorgiou,$^{b, }$\footnote{\tt papageorgiou.hep@gmail.com}  \\
Fuminobu Takahashi,$^{c, }$\footnote{\tt fumi@tohoku.ac.jp}  
and Takahiro Terada$^{b, }$\footnote{\tt takahiro.terada.hepc@gmail.com}  \\
}

\vspace{.4cm} 
{\small 
{\em $^a$Center for Gravitational Physics and Quantum Information, \\
Yukawa Institute for Theoretical Physics,
Kyoto University, 606-8502, Kyoto, Japan}\\
\vspace{.1cm} {\em $^b$Particle Theory  and Cosmology Group, Center for Theoretical Physics of the Universe, 
\\ Institute for Basic Science (IBS), Daejeon, 34126, Korea}\\
\vspace{.1cm} {\em $^c$Department of Physics, Tohoku University, Sendai, 980-8578, Miyagi, Japan}
}

\vspace{8 mm}

\begin{abstract}
Inflationary models, especially those with plateau-type potentials, are consistent with the cosmological data, but inflation itself does not resolve the initial singularity. This singularity is resolved, for example, by the idea of the quantum creation of the Universe from nothing such as the tunneling and no-boundary proposals. The simplest one predicts a closed Universe. Motivated by these facts, we investigate the classical dynamics of a closed universe with a plateau-type potential.
Depending on the initial inflaton field value, the universe can undergo a variety of events: an immediate big crunch, a bounce or cyclic phase, and inflation.  Although the non-inflationary solutions may appear irrelevant to our Universe, they can be turned into single or multiple bounces followed by inflation, taking into account the interactions necessary for the reheating of the Universe after inflation.  Thus, the dissipative mechanism in our setup explains both the graceful entry to and exit from inflation and gives us an indirect observational handle on the Universe just after creation. We also comment on the implications of these solutions for probabilistic interpretations of the wave function of the Universe. 
\end{abstract}

\end{center}
\vspace{1.2cm}
\end{titlepage}

\setcounter{footnote}{0}

\tableofcontents

%%%%%%%%%%%%%%%%%%%%%%%%%%%%%%%%%%%%%%%%%%%%%%%%%%%
\section{Introduction}

The observational data of the cosmic microwave background (CMB)~\cite{Planck:2018jri} and the large-scale structures~\cite{SDSS:2023tbz} are well explained by the $\Lambda$CDM model whose initial condition is given by the inflation paradigm. Namely, quantum fluctuations of the scalar field driving inflation called inflaton give the seeds of cosmological perturbations.  Moreover, inflation solves the puzzles of the standard big bang cosmology such as homogeneity, flatness, and absence of monopoles. In particular, the current data prefer inflation models with the inflaton potential having a very flat part (plateau) such as $R^2$ inflation~\cite{Starobinsky:1980te} and Higgs inflation~\cite{Bezrukov:2007ep}.

The inflation paradigm itself does not eliminate the initial big bang singularity as shown by the Borde-Guth-Vilenkin theorem~\cite{Borde:2001nh},\footnote{
Critical discussions on the theorem are given in Ref.~\cite{Lesnefsky:2022fen}.
} so some kind of past completion of the inflationary cosmology is necessary to understand the origin of the Universe. 
An elegant idea is the \textit{quantum creation of the Universe from nothing} that can be broadly classified as the \textit{tunneling proposal}~\cite{Vilenkin:1984wp, Vilenkin:1986cy, Vilenkin:1987kf} and the closely related \textit{no-boundary proposal}~\cite{Hawking:1981gb, Hartle:1983ai, Hawking:1983hj}. For example, in the latter picture, the initial singularity is just a coordinate singularity on smooth compact spacetime manifolds in the Euclidean path integral. 
The tunneling picture provides a singularity-free origin of the Universe by quantum tunneling, and related early proposals were given in the literature~\cite{Linde:1983cm,Linde:1983mx,Linde:1984ir,Rubakov:1984bh,Zeldovich:1984vk}. 
In such a singularity-free birth, the Universe has positive spatial curvature (see Refs.~\cite{Zeldovich:1984vk,Coule:1999wg, Linde:2004nz, Linde:2017pwt} for alternative spacetime topology), so it is worth exploring whether and how the Universe after creation can lead to inflation.

If the initial value of the inflaton field is on the plateau of the scalar potential, the sufficiently flat potential leads to inflation in the context of the quantum creation of the Universe. However, if the initial field value is close to the edge of the plateau, the inflation is naively too short, and the universe sooner or later collapses due to the large positive curvature, ending in the big crunch.

In the previous work by three of us~\cite{Matsui:2019ygj} (see also Ref.~\cite{Sloan:2019jyl}), we found exotic non-singular solutions for the homogeneous and isotropic universe in \ac{gr}: (1) a bounce (i.e., a transition from a contracting universe to an expanding universe) followed by inflation, and (2) cyclic solutions (i.e., repetition of contraction and expansion of the universe). Both require positive spatial curvature to realize a bounce, while the positive spatial curvature is also one of the simplest predictions of the space-time structure by the creation of the Universe from nothing.

For a universe with zero or negative spatial curvature, a bounce is ruled out by the \ac{nec} in \ac{gr}.  With positive spatial curvature, on the other hand, a bounce is possible when the curvature term is balanced by the total energy of the universe. This condition is generically hard to be satisfied in \ac{gr}, since the curvature term grows slowly compared to the kinetic energy of the scalar field, and the kinetic-energy-dominated collapsing universe soon reaches the big crunch. Nevertheless, we found that the sufficiently flat potential and initially large contributions of the positive spatial curvature, predicted by the creation from nothing, naturally lead to the bounce or cyclic universe~\cite{Matsui:2019ygj}.

While the non-singular solutions of type (1) can be easily connected to the standard inflationary cosmology, the non-singular solutions of type (2) may appear irrelevant to our Universe since the ratio of the maximum and the minimum of the scale factor is of order unity in our cyclic solutions~\cite{Matsui:2019ygj} (see also Appendix~\ref{sec:osc_avg}). In this paper, we show that the cyclic solutions can be connected to inflationary cosmology by taking into account interactions of the scalar field with other fields.  Such interactions are well motivated since, if the scalar field is to be identified with the inflaton at all, some interactions are necessary to reheat the Universe.

Beyond the homogeneous and isotropic approximation, the cyclic solutions are also affected by tachyonic instability, which is a built-in feature of the plateau-type potential, as we will show in this paper.  It is not surprising that an eternal cyclic solution is impossible.\footnote{ 
This has been known for a long time. See, e.g., Ref.~\cite{Tolman:1931fei}, although strictly speaking, the analyses there do not apply to our setup since the pressure was assumed to be non-negative.
} One may expect this on the ground of the second law of thermodynamics: the thermodynamic entropy never decreases, and entropy would increase during the nontrivial dynamics of the Universe in an interacting theory, invalidating the periodic solution.  However, it is nontrivial if the modified quasi-cyclic solutions lead to inflation since it can also lead to big crunch.\footnote{
For other works on quasi-cyclic Universe followed by inflation, see Refs.~\cite{Mulryne:2005ef, Biswas:2008kj, Biswas:2009fv, Yoshida:2019pgn}.
}  In fact, if the tachyonic instability is the dominant effect that affects the otherwise periodic solution, the scalar field fragments and relativistic modes are excited, ultimately leading to the kinetic-energy-dominated collapsing universe.  Thus, the question is if the interactions responsible for the reheating of the Universe can trigger also the beginning of inflation before the inflaton field fragments due to the tachyonic instability.

We answer this question positively in an accompanying letter~\cite{Matsui:2023wxm} and dub the mechanism ``\textit{dissipative genesis}'' of an inflationary universe. Here, we present more detailed analyses of the dissipative genesis. 
The organization of this paper is as follows.
In Sec.~\ref{sec:cyclic}, we define the model setup and present quasi-cyclic solutions.  The effect of the tachyonic instability is also discussed.  Sec.~\ref{sec:dissipation} is the main section, where we introduce dissipative effects and study how the quasi-cyclic solutions are modified. In particular, we see that the modified solutions can lead to inflation, whose prediction on the CMB observables is briefly summarized in Sec.~\ref{sec:CMB}.  The preheating and reheating of the Universe are discussed in Sec.~\ref{sec:reheating}.  Our analyses are at the classical level, but how the initial condition of the classical time evolution can be understood in quantum cosmology is discussed in Sec.~\ref{sec:creation}. 
We comment on the intriguing possibility that following the Hartle-Hawking no-boundary proposal, the Universe most likely experienced bounce(s) before inflation. Our conclusions are given in Sec.~\ref{sec:conclusions}.

We use the natural units $c = \hbar = M_\text{P}^{-2} (\equiv 8\pi G) = k_\text{B} = 1$ unless these symbols are explicitly denoted.

%%%%%%%%%%%%%%%%%%%%%%%%%%%%%%%%%%%%%%%%%%%%%%%%%%%
\section{Quasi-cyclic Universe \label{sec:cyclic}}

%%%%%%%%%%%%%%%%%%%%%%%%%%%%%%%%%%%%%%%%
\subsection{Setup}
In this section, we shall provide our setup.
We consider the following Lagrangian density:
\begin{align}
    \mathcal{L} = \frac{1}{2}R - \frac{1}{2}g^{\mu\nu}\partial_\mu \phi \partial_\nu \phi - V(\phi) + \mathcal{L}_\text{int}(\phi, \psi) + \mathcal{L}_\text{matter}(\psi), \label{full_theory}
\end{align}
where $R$ is the Ricci scalar, $g^{\mu\nu}$ is the inverse metric, $\phi$ is the canonically normalized scalar field of our interest, which can potentially lead to various dynamics including bounces and inflation, $V(\phi)$ is the scalar potential of $\phi$, $\psi$ collectively denotes other fields in the theory including the Standard Model degrees of freedom, $\mathcal{L}_\text{int}$ is the interaction between $\phi$ and $\psi$, and $\mathcal{L}_\text{matter}$ is the Lagrangian density for $\psi$.  In this section, we neglect $\psi$ and consider the dynamics of $\phi$ in the curved spacetime.

As an inflationary model consistent with the observational data, we consider a single-field slow-roll inflation model with the following plateau-type potential:
\begin{align}
    V(\phi) =& V_0 \tanh^n \left( \frac{\phi}{\phi_0} \right) + \Lambda, \label{potential}
\end{align}
where $V_0$ is the overall height of the potential, $\phi_0$ is the characteristic field value governing the width of the potential, and $\Lambda$ is the tiny cosmological constant. We will neglect $\Lambda$ in the following, but we have a short comment on its role in Sec.~\ref{sec:creation}. For simplicity, we set the power $n=2$, which allows some analytic studies (Appendix~\ref{sec:osc_avg}).  Identifying $\phi_0 \equiv \sqrt{6\alpha_\phi}$,\footnote{
To distinguish it from the gauge coupling $\alpha$, which we will introduce later, we express the parameter $\alpha$ of the $\alpha$-attractor by $\alpha_\phi$. 
} it is equivalent to T-model realizations of the $\alpha$-attractor models of inflation~\cite{Kallosh:2013hoa, Ferrara:2013rsa, Kallosh:2013yoa, Galante:2014ifa, Carrasco:2015pla}.  The $\alpha$-attractor can be viewed as a generalization of $R^2$ inflation and Higgs inflation, and the special value $\alpha_\phi=1$ reproduces the predictions of $R^2$ inflation and Higgs inflation. 

The only important feature of the above choice of potential is that it has a minimum and plateau(s). The details of the shape of the potential are not important, at least qualitatively, and other forms, e.g., E-model realizations of the $\alpha$-attractor, $V(\phi) = V_0 (1 - e^{-2\phi/\phi_0})^n$, and the inverse-hilltop potential, whose asymptotic form reads $V(\phi) \sim V_0 (1 - (\phi_0 / \phi)^n)$, can also be chosen.  The plateau region needs not to extend infinitely as long as it is sufficiently long. 
An advantage of the above form is that it allows some analytic calculations.

For simplicity, we assume that the Universe is created in a symmetric, i.e., homogeneous and isotropic configuration. We come back to this point in Sec.~\ref{sec:creation}. 
Before inflation, the spatial curvature cannot be neglected, 
so we take the Friedmann-Lema\^{i}tre-Robertson-Walker metric: 
\begin{align}
    \mathrm{d} s^2 = - \mathrm{d}t^2 + a(t)^2 \left(\frac{1}{1 - K r^2}\mathrm{d} r^2 + r^2 \left(\mathrm{d} \theta^2 + \sin^2 \theta \mathrm{d} \varphi^2 \right) \right),
\end{align}
where $a(t)$ is the scale factor, and $K$ is the spatial curvature, which is predicted to be positive and nearly maximal in the context of the creation of the Universe from nothing. We also take the homogeneous scalar field $\phi(t)$ for the moment.  

In the calculation of the wave function of the Universe, it is often assumed that the scalar field is on a sufficiently flat part of the scalar potential~\cite{Vilenkin:1987kf},\footnote{
See Refs.~\cite{Matsui:2020tyd, Janssen:2020pii} for more detailed discussions.
}
\begin{align}
\left| V' \right| \ll \max \left [ |V|, \, \frac{K}{a^2} \right ], \label{flatness_condition}
\end{align}
where a prime denotes differentiation with respect to its argument $V'(\phi) = \mathrm{d} V(\phi) / \mathrm{d} \phi$,  
so that the derivative of the wave function with respect to $\phi$ can be neglected.  In the quantum creation regime, i.e., in the classically forbidden region, the second term on the right-hand side is dominant, so $|V'| \ll |V|$ is a \emph{sufficient} condition to use the results of quantum cosmology calculations where $V\approx \text{const.}$ is assumed. This is essentially the same as the slow-roll condition for inflation.  

 Provided that the flatness condition~\eqref{flatness_condition} is satisfied, the initial condition for the classical time evolution after quantum creation is given by~\cite{Vilenkin:1987kf} 
\begin{align}
    \dot{a} =& 0 & \text{and}&  &  
    \dot{\phi} =& 0, \label{initial_condition}
\end{align}
where the overdot denotes the time derivative, $a = \sqrt{3K/V(\phi)}$ is determined by the Friedmann equation (see below), and the value of the homogeneous field  $\phi$ is arbitrary.  The probability distribution of the initial field value $\phi$ is discussed in Sec.~\ref{sec:creation}. 
As mentioned in the Introduction, we are primarily interested in the cases where the initial field value $\phi$ is close to the boundary of the flat region of the potential. Otherwise, if it is deep in the plateau region, long-lasting inflation is easily realized and there are no issues to discuss for our purposes. On the other hand, if the initial value completely violates the flatness condition, we cannot rely on, e.g., Eq.~\eqref{initial_condition}. 

The time evolution is governed by the Friedmann equation and the equation of motion for $\phi$, 
\begin{align}
    &H^2 = \frac{1}{3} \left(\frac{1}{2} \dot{\phi}^2 + V(\phi) - \frac{3K}{a^2} \right), \\
    &\ddot{\phi}+ 3H\dot{\phi} + V' = 0,
\end{align}
where $H(t) \equiv \dot{a}/a$ is the Hubble parameter. 

We will take into account the development of inhomogeneity in $\phi$ later in this section, but we neglect inhomogeneity in the metric.  This is because we focus on the sub-Planckian regime of energy density, so the gravitational backreaction effect is expected to be small.  Anyway, we are ultimately interested in the dynamics before the inhomogeneity in $\phi$ would become important, so the inhomogeneity in the gravitational degrees of freedom can be safely neglected. 

%%%%%%%%%%%%%%%%%%%%%%%%%%%%%%%%%%%%%%%%
\subsection{Cyclic universe solutions}

When the initial field value $\phi$ is close to the edge of the plateau, the slow-roll condition is soon violated, so the accelerated expansion quickly ends before the spatial curvature is significantly diluted. One of the typical fates of such a closed universe is to collapse into a big crunch, $a(t) \to 0$. This is true especially when $\phi_0$ is Planckian or super-Planckian.  However, there are also cyclic solutions.  

%%%%
\begin{figure}[htbp]
    \centering
    \subcaptionbox{$\phi(t) \quad (\phi_0 =\sqrt{6} \times 10^{-2}) $}{
    \includegraphics[width=0.48\columnwidth]{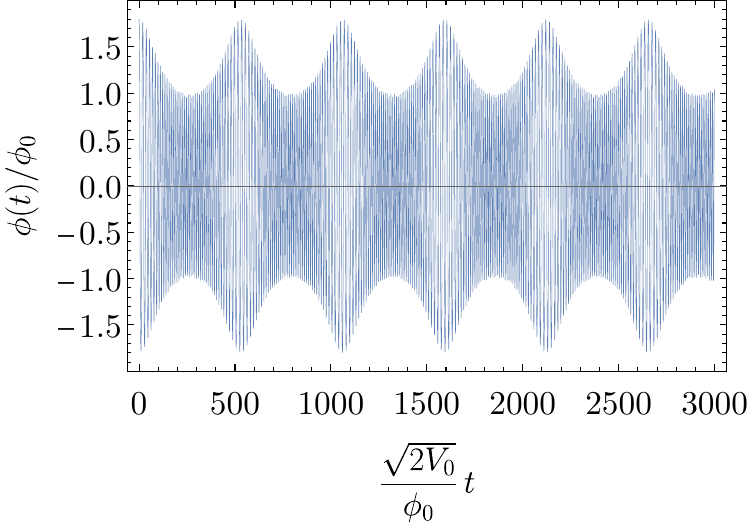}
    }~\subcaptionbox{$a(t) \quad (\phi_0 =\sqrt{6} \times 10^{-2}) $}{
    \includegraphics[width=0.48\columnwidth]{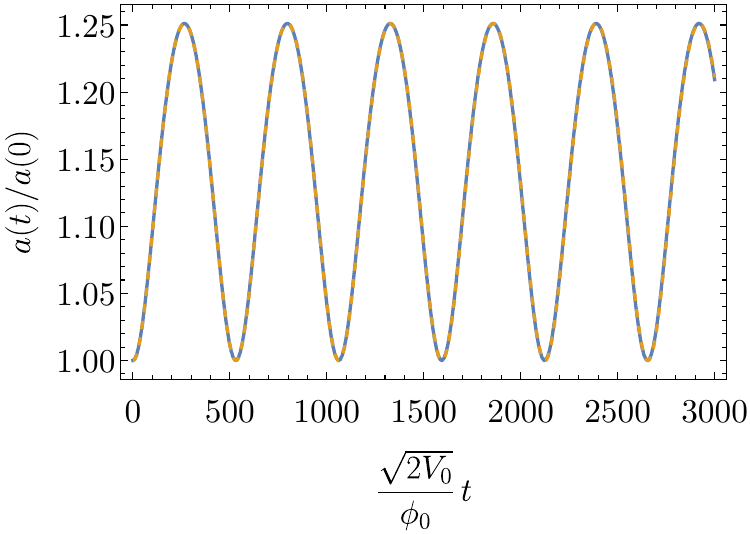}
    }
    \vskip 2mm
    \subcaptionbox{$\phi(t) \quad (\phi_0 = \sqrt{6} \times 10^{-1}) $}{
    \includegraphics[width=0.48\columnwidth]{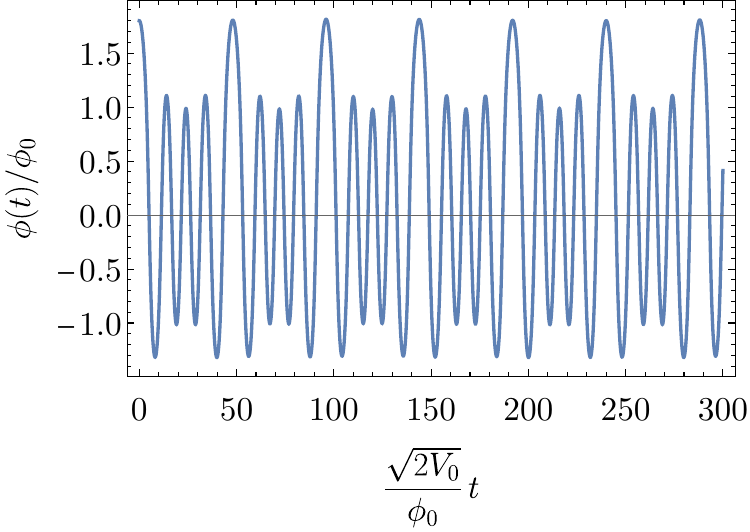}
    }~\subcaptionbox{$a(t) \quad (\phi_0 = \sqrt{6} \times 10^{-1}) $}{
    \includegraphics[width=0.48\columnwidth]{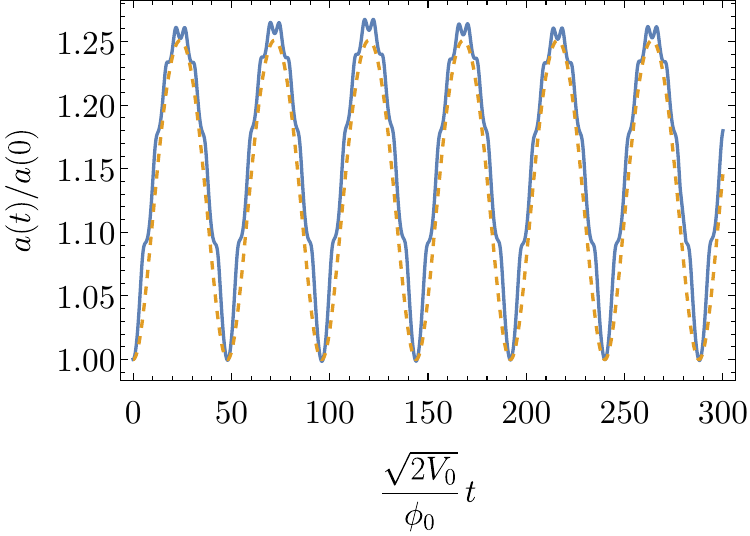}
    }
    \caption{Cyclic universe solutions for the scalar field $\phi(t)$ (left panels) and the scale factor $a(t)$ (right panels).  $\phi_0 = \sqrt{6} \times 10^{-2}$ and $\sqrt{6} \times 10^{-1}$ for the top and bottom rows, respectively.  The solid and dashed lines for $a(t)$ (right panels) show an exact and an oscillation-averaged solution, respectively. For the bottom right panel, the oscillation-averaged solution underestimates the frequency of the oscillation, so the correction factor $1.107$ has been introduced by hand.  The initial amplitude is $\phi_\text{ini}/\phi_0 = 1.8$ for all of the panels.  The effects of any dissipation including tachyonic instability are neglected. 
    }
    \label{fig:cyclic}
\end{figure}
%%%%

Examples of the cyclic solutions are shown in Fig.~\ref{fig:cyclic}. 
These kinds of cyclic solutions in our setup were discovered in Ref.~\cite{Matsui:2019ygj}.  The interpretation given there for the mechanism of cycles is as follows. Suppose that $\phi$ is oscillating around the minimum and $\phi_0 \ll 1$ so that the dynamics of $\phi$ is much faster than that of $a$.  In this case, we can take the oscillation average with respect to the motion of $\phi$.  Now, $\phi$ can be viewed as a cosmological fluid with an effective equation-of-state parameter $w_\text{avg} = \langle P \rangle_\text{osc}/\rho $, where $P$ and $\rho$ are the pressure and the energy density, respectively, and $\langle \cdot \rangle_\text{osc}$ denotes the oscillation average.  The dependence of $w_\text{avg}$ on the slowly varying amplitude $\phi_\text{amp}$ is derived in Appendix~\ref{sec:osc_avg} and shown in Fig.~\ref{fig:w_avg} there. 
During an expanding phase of the Universe, the amplitude of $\phi$ decreases due to the Hubble friction. When the amplitude becomes sufficiently small, $w_\text{avg} > - 1/3$.  At some point, the effective energy density of the spatial curvature $\rho_K \equiv - 3K/a^2$, which has an effective equation-of-state parameter $w=-1/3$ (i.e., $\rho_K \propto a^{-2}$), becomes equal to the positive energy density of $\phi$, and $H$ vanishes.  Since the second derivative of the scale factor is given by 
\begin{align}
    \frac{\ddot{a}}{a} =& - \frac{1}{6}(\rho + 3 P) \nonumber \\
    =& - \frac{\rho}{6}(1 + 3 w),
\end{align}
the expanding Universe \textit{turns around} into a contracting Universe. 
During a contracting ($H<0$) phase of the Universe, the amplitude of $\phi$ \emph{increases} due to the Hubble \emph{anti}-friction. When the amplitude becomes sufficiently large, $w_\text{avg} < - 1/3$. At some point, the effective energy density of the spatial curvature becomes equal to the positive energy density of $\phi$, and $H$ vanishes.  This time, the sign of $\ddot{a}$ is positive, and the contracting Universe \textit{bounces} back to an expanding Universe. This feedback mechanism persists, and the solution becomes cyclic. This explanation boils down to Eq.~\eqref{Friedmann_eq_avg2} in Appendix~\ref{sec:osc_avg} and corresponds to the orange dashed lines in Fig.~\ref{fig:cyclic}.  For a sufficiently small $\phi_0 \ll 1$ (the top row), it excellently matches with the exact solution (the solid blue line).

For a larger $\phi_0$, the oscillation-average approximation is not good enough.  For example, it underestimates the frequency of the oscillations of $a(t)$ in Fig.~\ref{fig:cyclic}~(d), so the correction factor 1.107 has been introduced by hand. Moreover, when $\phi_0 \sim \mathcal{O}(1)$, even a  single bounce requires fine-tuning of either $\phi_0$ or the initial value of $\phi$. Later, we will be interested in a regime [cf.~bottom panels in Fig.~\ref{fig:cyclic}] in which an $\mathcal{O}(10\%)$ tuning of the initial value of $\phi$ for a given $\phi_0$ is required for a reason related to the tachyonic instability, which we will discuss next.

%%%%%%%%%%%%%%%%%%%%%%%%%%%%%%%%%%%%%%%%
\subsection{Tachyonic instability \label{ssec:tachyonic}}
So far, our consideration was limited to the homogeneous level.  However, Fourier modes with finite wavenumber $k$ are generated (amplified) by the self-interactions of $\phi$. For our potential, tachyonic instability and parametric resonance can occur.  A tachyonic part of the potential is generally expected for a plateau-type potential between the inflection point and the plateau.  Parametric resonance occurs because the effective mass of the finite momentum modes of $\phi$ has a non-adiabatic time dependence around when the zero mode of $\phi$ passes the minimum of the potential.  These effects are well studied in the context of preheating after inflation.  Since we are interested in the solutions with oscillations of $\phi$, these `non-perturbative' particle production effects are also important in our setup.  In particular, the tachyonic instability is strong for the shape of our potential.

There are also differences from preheating after inflation.  First, $H$ oscillates around 0 in quasi-cyclic solutions, and its amplitude is suppressed by the cancellation between positive $\rho$ and negative $\rho_K$.  Because of this, it is expected to be closer to the flat spacetime result.  In particular, compared to the standard preheating, it is harder for unstable modes to exit the instability band because there is no net redshift or blueshift due to Hubble expansion or contraction.  The instability grows until the nonlinearity and the backreaction to the background become non-negligible. 

Second, the initial condition is different compared to the standard preheating.  The initial condition for the background is given in Eq.~\eqref{initial_condition}. 
We assume that the initial condition for the finite momentum modes is a  vacuum-like state so that we can use the same formula as in the standard preheating.  This is enough for our purpose since the tachyonic instability grows exponentially as a function of time, so the instability time scale is only logarithmically sensitive to the initial condition.

With these similarities and differences in mind, we can use the results of Ref.~\cite{Tomberg:2021bll}\footnote{
See Refs.~\cite{Lozanov:2017hjm, Turzynski:2018zup} for complementary analyses and Refs.~\cite{Krajewski:2018moi, Iarygina:2018kee, Iarygina:2020dwe} for multifield generalizations.
}  with appropriate modifications. 
The energy density of perturbations can be obtained by the saddle-point approximation~\cite{Tomberg:2021bll}\footnote{ 
In deriving this formula, the mass-term contribution (local curvature of the potential) to the dispersion relation for each momentum mode $k$ is neglected.  This is well justified for large momenta and will affect small-momentum modes by a factor of $\mathcal{O}(1)$ coefficient.  In the simulation by \CosmoLattice{ }described below, the default setting we use also neglects the mass contribution when it is tachyonic.
}  
\begin{align}
    \delta \rho \approx \frac{(k_\text{peak}/a)^4}{4\pi^{3/2} \sqrt{2 \mu_\text{peak} t}} e^{2\mu_\text{peak} t}.
\end{align}
where $k_\text{peak}/a \approx \pi/\Thalf$ is the wavenumber of the most unstable mode with $\Thalf$ the half-period of the oscillations of $\phi$, $\mu_\text{peak} \approx \mathcal{O}(1) /\Thalf$ is the growth exponent of the most unstable mode.  See Ref.~\cite{Tomberg:2021bll} for numerical values of the coefficients of these quantities.  

%%%%
\begin{figure}[htbp]
    \centering
    \subcaptionbox{volume average of $\phi(t,\bm{x})$ \label{sfig:CL_phi}}{
    \includegraphics[width=0.48\columnwidth]{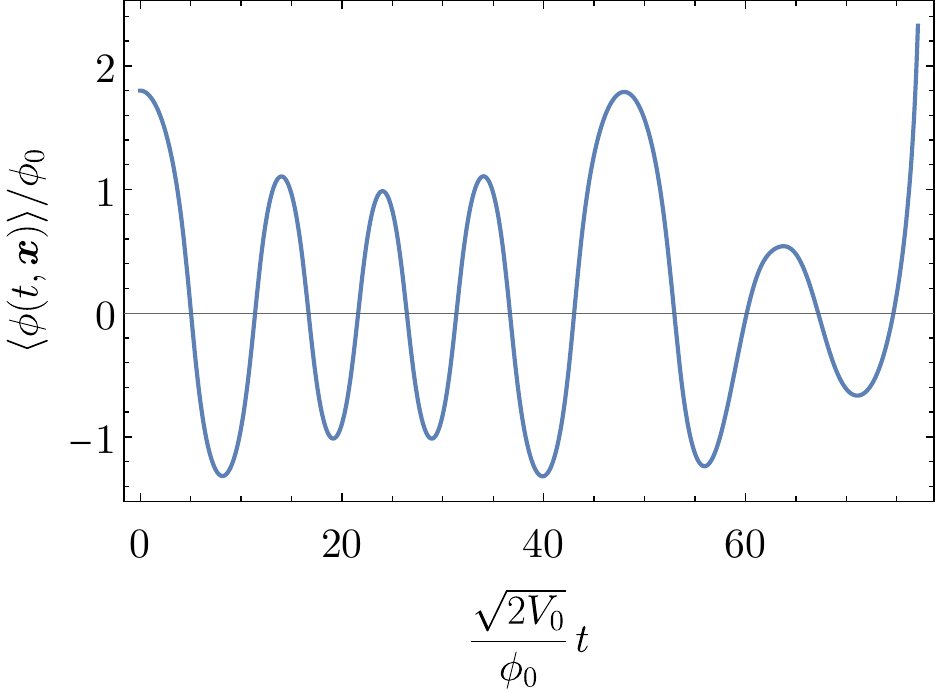}
    }~\subcaptionbox{$a(t)$ \label{sfig:CL_a}}{
    \includegraphics[width=0.48\columnwidth]{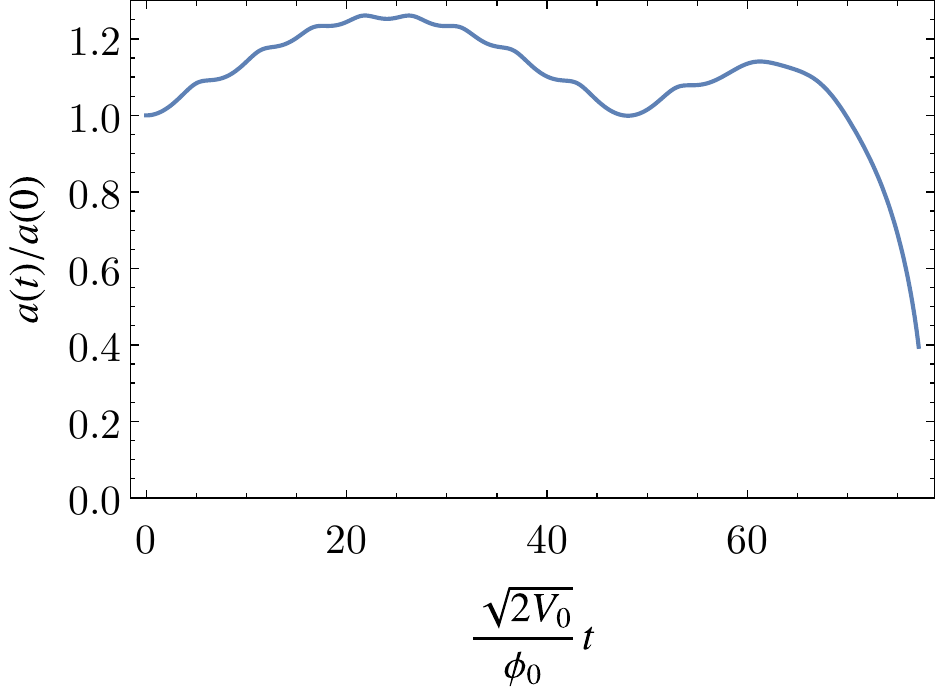}
    }
    \vskip 2mm
    \subcaptionbox{volume average of energy density components \label{sfig:CL_rho}}{
    \includegraphics[width=0.7\columnwidth]{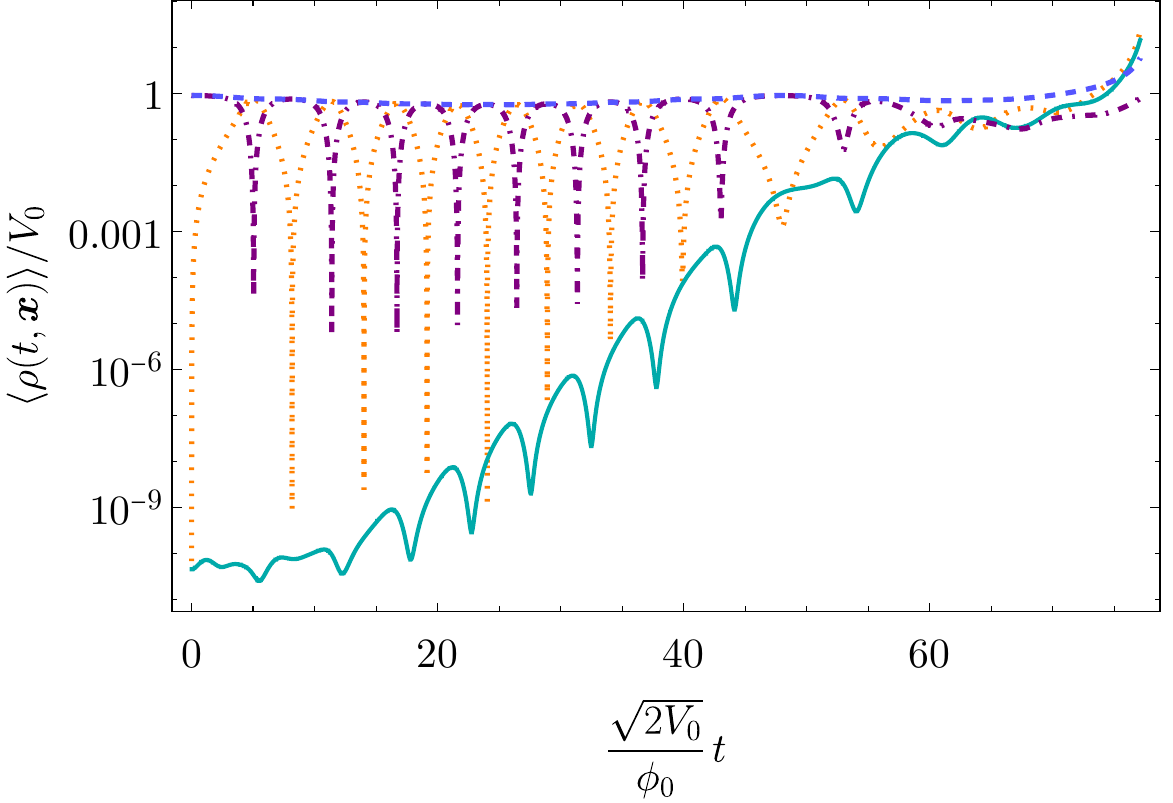}
    }
    
    \caption{Classical lattice simulation of the quasi-cyclic Universe by \CosmoLattice. \eqref{sfig:CL_phi} Time evolution of the volume average of $\phi(t, \bm x)$. \eqref{sfig:CL_a} Time evolution of $a(t)$, approaching the big crunch.  \eqref{sfig:CL_rho} Volume averages of energy components are plotted as a function of time.  The dark cyan solid line shows the gradient energy growing due to the tachyonic instability. It reaches $\mathcal{O}(1\%)$ of $V_0$ after the bounce at around $(\sqrt{2V_0}/\phi_0) \, t \approx 48$. The light blue dashed line shows the absolute value of the curvature contribution $|\rho_K|$, which cancels a substantial fraction of the positive energy contributions to the right-hand side of the Friedmann equation.  The purple dot-dashed line shows the potential energy. The orange dotted line shows the kinetic energy, which finally dominates the energy density and leads to the big crunch. The parameters are the same as those for the bottom row of Fig.~\ref{fig:cyclic} and $V_0 = (2.6 \times 10^{15}\,\text{GeV})^4$ fits the CMB normalization (see Sec.~\ref{sec:CMB}).}
    \label{fig:lattice}
\end{figure}
%%%%

When $\delta \rho$ becomes comparable to the background energy density $\rho$, i.e., $\delta \rho = c \rho$ with $c$ being an $\mathcal{O}(1)$ constant, the cyclic solutions found in the previous subsection are significantly modified because of the backreaction. Since the gradient energy becomes comparable to the kinetic and potential energies, we call this moment the fragmentation time $t_\text{frag}$. This is estimated as follows:
\begin{align}
    t_\text{frag} \approx & \frac{1}{2 \mu_\text{peak}} \left( \ln \left( \frac{4\pi^{3/2}  c \rho}{(k_\text{peak}/a)^4} \right) + \frac{1}{2} \ln   \ln \left( \frac{4\pi^{3/2}  c \rho}{(k_\text{peak}/a)^4} \right)  \right). \label{fragmentation_time}
\end{align}
For numerical evaluation, Eq.~(3.7) and Figure 4 of Ref.~\cite{Tomberg:2021bll} are useful. For example, for $1.31 \lesssim \phi_\text{amp}/\phi_0 \lesssim 3$, these show $1.5 \lesssim \mu_\text{peak} \Thalf \lesssim 3.93$ and $0.8 \lesssim k_\text{peak}\Thalf/(\pi a) \lesssim 1.13$. 
If we substitute $\mu_\text{peak} \approx 2.5 /\Thalf$, $k_\text{peak}/a \approx \pi /\Thalf$, $\rho \approx 0.93 V_0$ (corresponding to $\phi_\text{amp}/\phi_0 \approx 2$), $\phi_0 = \sqrt{6}\times 10^{-1}$, $V_0 = 5 \times 10^{-12}$, and $c = 0.1$, we have $t_\text{frag} \approx 2.7 \times 2 \Thalf$. That is, within a few oscillations of $\phi$, the background solution receives a significant backreaction.  $t_\text{frag}$ in units of the characteristic time scale $1/\sqrt{V_0}$ can be increased, e.g., by lowering the energy scale $V_0$. However, its dependence is only logarithmic.  Thus, the tachyonic instability is rather severe. 

This analytic result is consistent with numerical simulations.  We have modified \CosmoLattice~\cite{Figueroa:2020rrl, Figueroa:2021yhd}, a public tool for the classical lattice simulation, to include the spatial curvature.  Precisely speaking, we should modify the spatial lattice and its boundary conditions from torus $\mathbb{T}^3$ to sphere $\mathbb{S}^3$, but it is beyond the scope of our analyses.  For the details of the calculations using \CosmoLattice{}, see Appendix~\ref{sec:lattice}.

An example of the simulated result is shown in Fig.~\ref{fig:lattice}. We see that after the fragmentation time, the kinetic energy eventually dominates the dynamics since $\rho_\text{kin}\equiv \frac{1}{2}\dot{\phi}^2 \propto a^{-6}$ ($w_\text{kin} = 1$), and the universe collapses into the big crunch.  

%%%%%%%%%%%%%%%%%%
\begin{figure}[htbp!] 
\begin{center}
\includegraphics[width=0.49 \columnwidth]{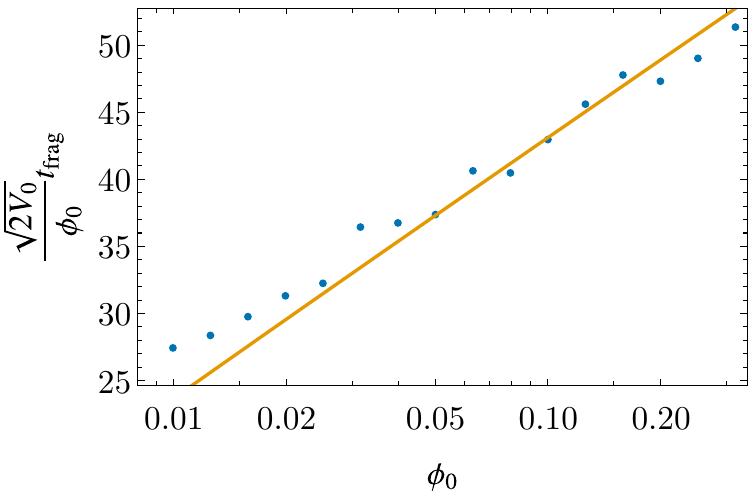}
\includegraphics[width=0.49 \columnwidth]{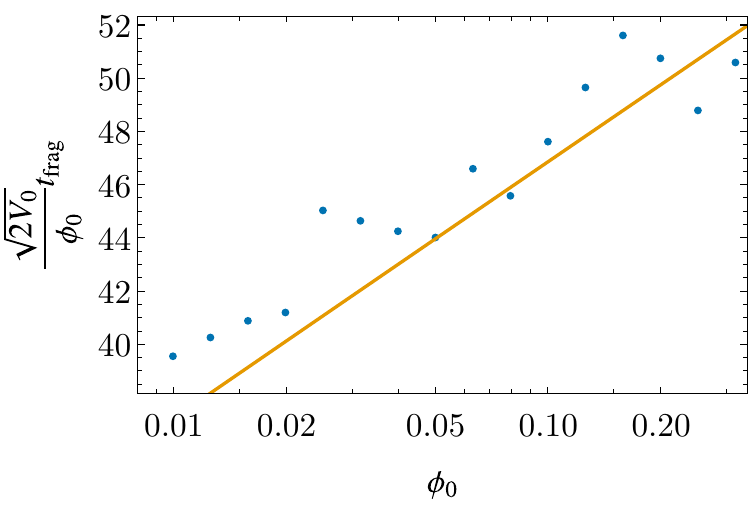}
\includegraphics[width=0.49 \columnwidth]{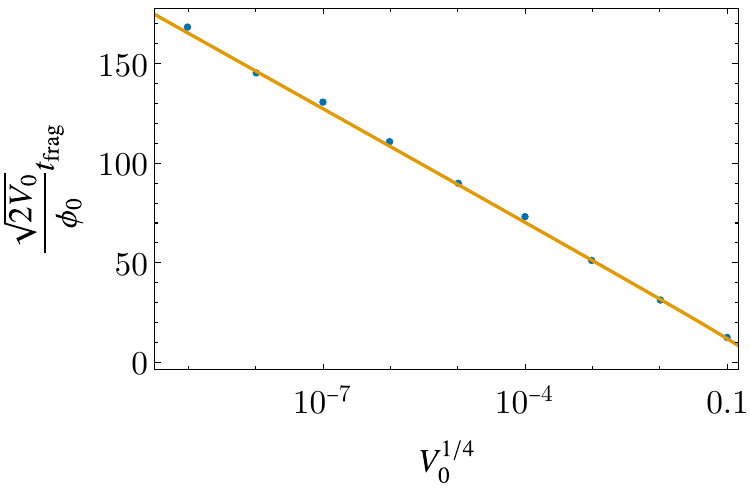}
\includegraphics[width=0.49 \columnwidth]{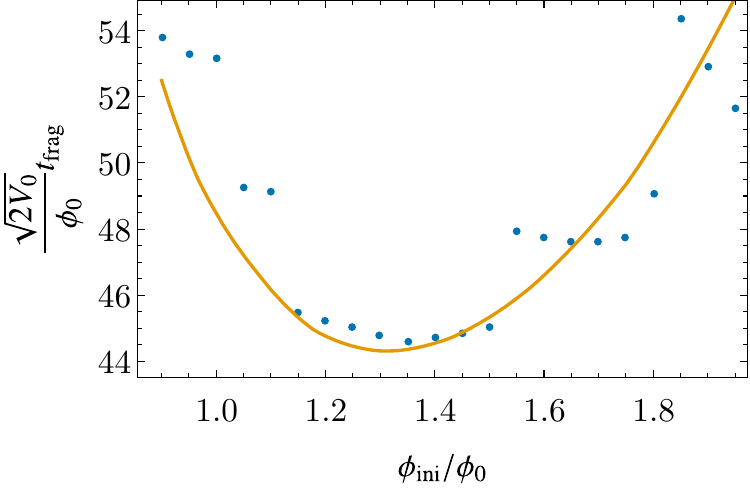}
\caption{Dependence of the fragmentation time on parameters. Top left:\, $\phi_0$ dependence with fixed $V_0^{1/4}=2.6 \times 10^{15} \, \text{GeV}$ and $\phi_\text{ini}/\phi_0 = 1.8$. Top right:\, $\phi_0$ dependence with fixed $\phi_0/\sqrt{V_0}=2.1 \times 10^5$ to fit the CMB normalization and $\phi_\text{ini}/\phi_0 = 1.8$. Bottom left:\, $V_0$ dependence with fixed $\phi_0 = \sqrt{0.06}$ and $\phi_\text{ini}/\phi_0 = 1.8$. Bottom right:\,  $\phi_\text{ini}/\phi_0$ dependence with fixed $\phi_0=\sqrt{0.06}$ and $V_0^{1/4}=2.6 \times 10^{15}\,\text{GeV}$. The orange lines are based on the analytic estimate~\eqref{fragmentation_time} with numerical values $\mu_\text{peak}$ and $k_\text{peak}$ from Ref.~\cite{Tomberg:2021bll}.}
\label{fig:t_frag_dependence}
\end{center}
\end{figure}
%%%%%%%%%%%%%%%%%%
%%%%%%%%%%%%%%%%%%
\begin{figure}[htbp!] 
\begin{center}
\includegraphics[width=0.49 \columnwidth]{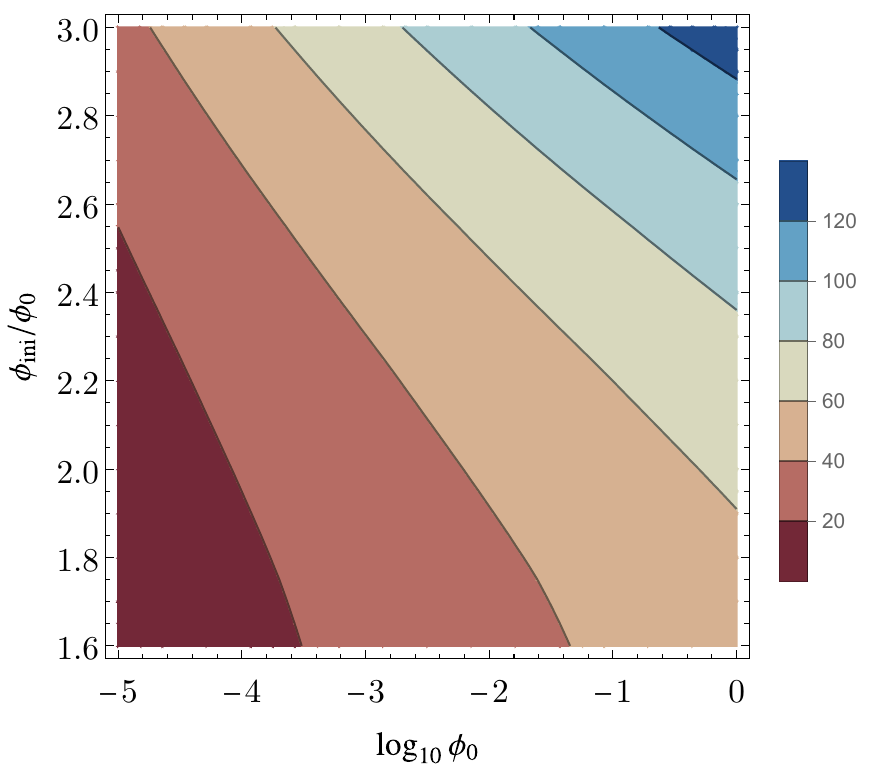}
\includegraphics[width=0.49 \columnwidth]{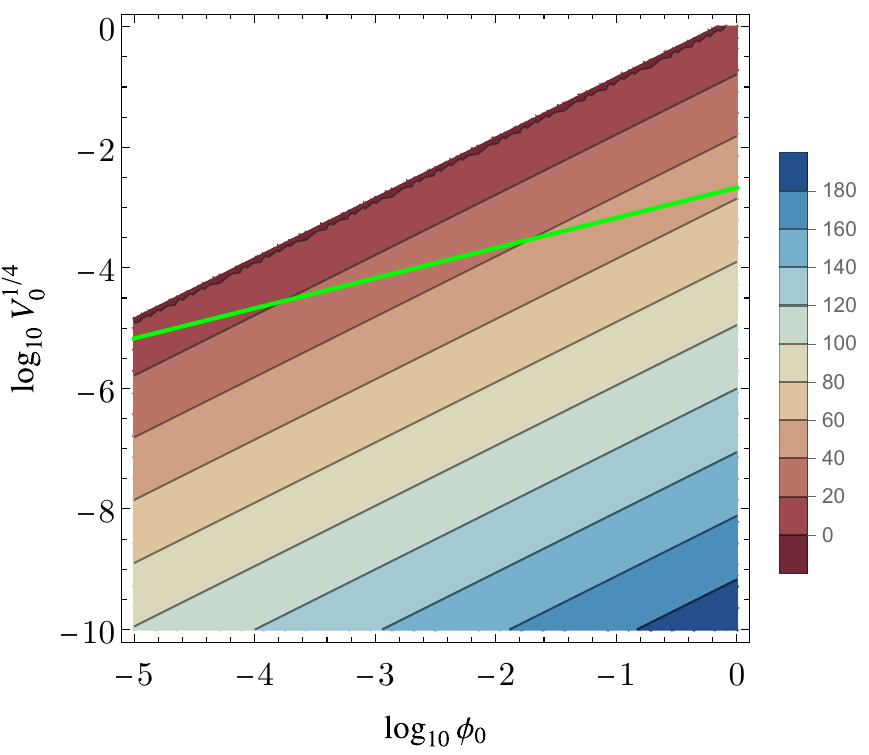}
\caption{The contour plots of $\frac{\sqrt{2V_0}}{\phi_0} t_\text{frag}$. Left:\, $\phi_0/\sqrt{V_0}=2.1 \times 10^5$ is kept fixed. Right:\, $\phi_\text{ini}/\phi_0=1.8$ is kept fixed. The light green line shows where the CMB normalization is fit.}
\label{fig:t_frag_contour}
\end{center}
\end{figure}
%%%%%%%%%%%%%%%%%%

For later convenience, we compare the fragmentation time formula~\eqref{fragmentation_time} with the corresponding numerical results of \CosmoLattice{} and validate the former since the lattice calculation is time-consuming. 
The procedure for comparing the fragmentation time is as follows. Since we are exploring various parameters, the universe may experience short inflation at the beginning and the energy density $\rho$ may be diluted.  Therefore, we compare the gradient energy with the time-dependent energy density $\rho(t)$, which is the sum of the kinetic, gradient, and potential energy density, with \CosmoLattice. We used $\rho_\text{grad}(t_\text{frag}) = 0.01 \rho(t_\text{frag})$ to extract the fragmentation time from the \CosmoLattice{} simulation.  This is compared in Fig.~\ref{fig:t_frag_dependence} with the more naive analytic estimate in Eq.~\eqref{fragmentation_time} with the substitution of the initial energy density $\rho$.  The numerical lattice results shown by dots and the analytic estimates shown by lines are consistent with each other. The fluctuations seen for the blue dots are caused by the oscillating features of $\rho_\text{grad}(t)$ as a function of time (see Fig.~\ref{sfig:CL_rho}). Depending on which peak hits the threshold, the fragmentation time changes discretely.  For the analytic lines, we plugged in numerical data of $\mu_\text{peak}$ and $k_\text{peak}$ from Fig.~4 of Ref.~\cite{Tomberg:2021bll}.  Encouraged by the agreement of these results, we use the (semi)analytic formula~\eqref{fragmentation_time} in the following to compute $t_\text{frag}$.  The contour plots of $t_\text{frag}$ in units of $\phi_0/\sqrt{2V_0}$ are shown in Fig.~\ref{fig:t_frag_contour}.

In summary, the cyclic solutions become quasi-cyclic solutions ending with the big crunch due to the tachyonic instability built in our plateau-type potential. 
For a more realistic Universe, however, we have to introduce other fields and interactions among them.  This drastically changes the fate of the Universe. 

%%%%%%%%%%%%%%%%%%%%%%%%%%%%%%%%%%%%%%%%%%%%%%%%%%%
\section{Dissipative effects and graceful entry to inflation \label{sec:dissipation}}

%%%%%%%%%%%%%%%%%%%%%%%%%%%%%%%%%%%%%%%%
\subsection{General discussion \label{ssec:dissipation_general}}

Let us restore $\mathcal{L}_\text{int}$ and $\mathcal{L}_\text{matter}$ in Eq.~\eqref{full_theory}. Interactions enable $\phi$ to decay, annihilate, or scatter, etc. These effects lead to the dissipation of the energy density of $\phi$ and increase the energy density of daughter particles and the entropy of the system.  In contrast to the tachyonic instability, these effects do not exponentially enhance the inhomogeneity. They are effectively homogeneous on the scales larger than the mean free path of particles, so the effects are qualitatively different from the tachyonic instability discussed in the previous section.   They affect the periodicity of the quasi-cyclic solutions at the homogeneous level.  
 In this section, we demonstrate that the interactions can turn the quasi-cyclic solutions into inflationary solutions.

If one studies such a system in a concrete way and in detail, one should specify $\mathcal{L}_\text{int}$ and $\mathcal{L}_\text{matter}$ and examine the equations of motion of the one-point correlation function $\langle \phi \rangle$ along with the scale factor $a$ taking into account quantum and/or thermal corrections. 
Since our primary purpose in this paper is to point out the existence of solutions with their qualitatively new behavior and to demonstrate our basic ideas without model-specific assumptions, however, we take a more phenomenological approach based on the Boltzmann equations introduced below.  We also study a concrete model in Sec.~\ref{sec:thermal-dissipation}.

Suppose that the energy of $\phi$ is dissipated into relativistic degrees of freedom (radiation). Generalization to non-relativistic degrees of freedom is straightforward.  
We suppose that the effects of the dissipation can be described by the following equations of motion: 
\begin{align}
    & \ddot{\phi} + (3 H + \Gamma ) \dot{\phi} + V' = 0, \label{eq:scalareom}\\
    & \dot{\rho}_\text{r} + 4 H \rho_\text{r} =  \Gamma \dot{\phi}^2,\label{eq:radiation} \\
    & H^2 = \frac{1}{3} \left( \rho_\phi + \rho_\text{r} \right) - \frac{K}{a^2}, \label{eq:friedman}
\end{align}
where $\Gamma$ is the dissipation rate, and $\rho_\phi$ and $\rho_\text{r}$ are the energy density of $\phi$ and radiation, respectively. 
We assume that there is no or a negligible amount of radiation just after the creation of the Universe but we will also briefly comment on the possibility that the Universe is created with a finite amount of radiation in Sec.~\ref{sec:creation}.

The dissipation rate $\Gamma$ in Eqs.~\eqref{eq:scalareom} and \eqref{eq:radiation} can have various field-theoretic origins, depending on which different levels of assumptions and approximations are needed to derive these equations. 
For example, when $\phi$ is strongly dissipated and adiabatically moves in the field space, the perturbative approach based on the in-in (Schwinger-Keldysh) formalism is useful.  Such formalisms to derive $\Gamma$ and general expressions of $\Gamma$ with or without finite temperature effects can be found in Refs.~\cite{Morikawa:1984dz, Morikawa:1986rp, Gleiser:1993ea, Berera:1998gx}. In particular, the above set of equations is the same as those derived in the context of warm inflation~\cite{Berera:1995ie, Berera:1998gx}. The dissipation rate can play an important role in the context of naturally explaining the initial conditions of inflation~\cite{Berera:2000xz, Bastero-Gil:2016mrl}. Also, it can affect the dynamics of a phase transition~\cite{Bartrum:2014fla} and have an implication on baryon asymmetry of the Universe~\cite{Bartrum:2014fla}. 

When the field oscillates around the minimum of the potential, on the other hand, parametric resonance can occur~\cite{Traschen:1990sw, Kofman:1997yn}. It is non-adiabatic dynamics.  In this case, a semiclassical analysis of particle production in the presence of a classical background of $\phi(t)$ is more useful than the perturbative Schwinger-Keldysh approach.  Averaging over the oscillation period of $\phi$, one can define the particle production rate $\Gamma_\text{prod}$ ($\dot{n}_\chi = \Gamma_\text{prod} n_\chi$  with $n_{\chi}$ being the number density of the produced particles $\chi$).  At the initial moment when there are no produced particles, the rate in the narrow resonance regime coincides with the perturbative decay rate $\Gamma_\text{dec}$ of $\phi$ ($\dot n_\chi = \Gamma_\text{dec} n_\phi$ with $n_{\phi}$ being the number density of $\phi$) at least when the potential around the minimum can be approximated as quadratic (see, e.g., Ref.~\cite{Braden:2010wd} and also Appendix A of Ref.~\cite{Ema:2016hlw}).\footnote{
When the Fourier component of the effective mass of the produced particles is dominated by higher frequencies than the fundamental frequency of the driving force, it corresponds to annihilation of $\phi$ quanta.  See, e.g., Ref.~\cite{Braden:2010wd} and Appendix B of Ref.~\cite{Ema:2016oxl} for this point.
}  The role of $\Gamma = \Gamma_\text{dec}$ in Eq.~\eqref{eq:scalareom} in this context is widely known as a phenomenological description of the damping effect originating from the imaginary part of the full propagator, which is nothing but the decay rate of $\phi$ due to the optical theorem, at least in the flat spacetime and with the vacuum state (for daughter particles)~\cite{Abbott:1982hn} (see also Ref.~\cite{Kofman:1997yn}).  

In our setup, $\phi$ oscillates around the minimum, but $\phi$ also moves slowly on a plateau.  This means that multiple sources of dissipation may work in different regimes of dynamics of $\phi$.

Before discussing an explicit model in Sec.~\ref{sec:thermal-dissipation}, let us discuss the above-simplified setup with a constant $\Gamma$ to demonstrate that the basic mechanism does not depend on details of the models or the form of the dissipation term.  The choice of a constant $\Gamma$ is for simplicity and to avoid introducing many parameters to describe possible time (or field) dependence of $\Gamma$. 
Of course, the realization of a constant $\Gamma$ will be nontrivial.  For example, with the above caveats, we may interpret $\Gamma$ as the decay rate.  However, the decay rate will realistically depend on the field value of $\phi$ because of the kinematic suppression. The effective mass of $\phi$ is suppressed on the plateau, so it will not decay there.  Motivated by this fact, we have studied a few examples of the field dependence $\Gamma(\phi)$ such as a smoothed top hat shape and a function mimicking kinematic suppression $ \sim \sqrt{1 - (\phi/\phi_0)^2}$.  We found that they are qualitatively similar to the case with the constant $\Gamma$.  This means that the dissipation mechanism operating only in the non-adiabatic (oscillatory) regime is sufficient to qualitatively reproduce the result of the constant $\Gamma$ case, for which the results are presented below.

To simplify the discussion, let us pretend for a moment that there is no tachyonic instability. After finding an interesting event that modifies the cyclic solutions and the time scale that occurs, we will compare the time scale of the event and $t_\text{frag}$.  If the former is sufficiently short, the tachyonic instability does not play a major role. 

%%%%
\begin{figure}[htbp]
    \centering
    \subcaptionbox{$\phi(t)$ \label{sfig:dsp_phi}}{
    \includegraphics[width=0.48\columnwidth]{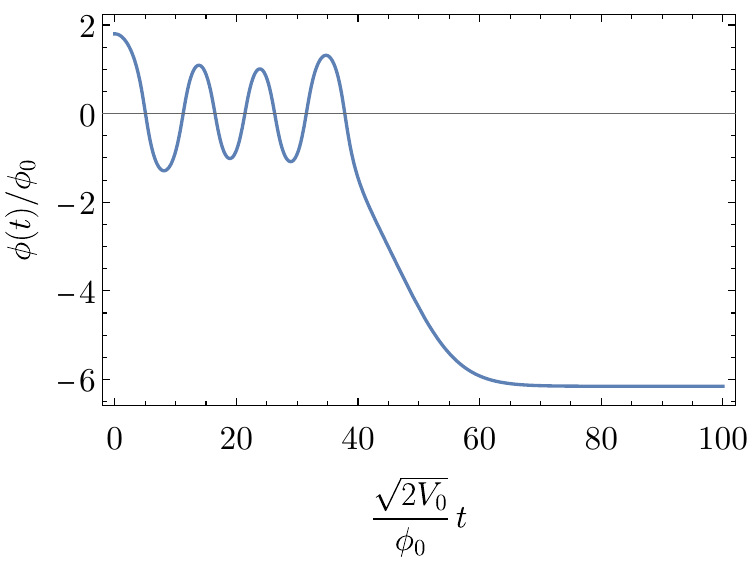}
    }~\subcaptionbox{$a(t)$ (log-plot) \label{sfig:dsp_a}}{
    \includegraphics[width=0.48\columnwidth]{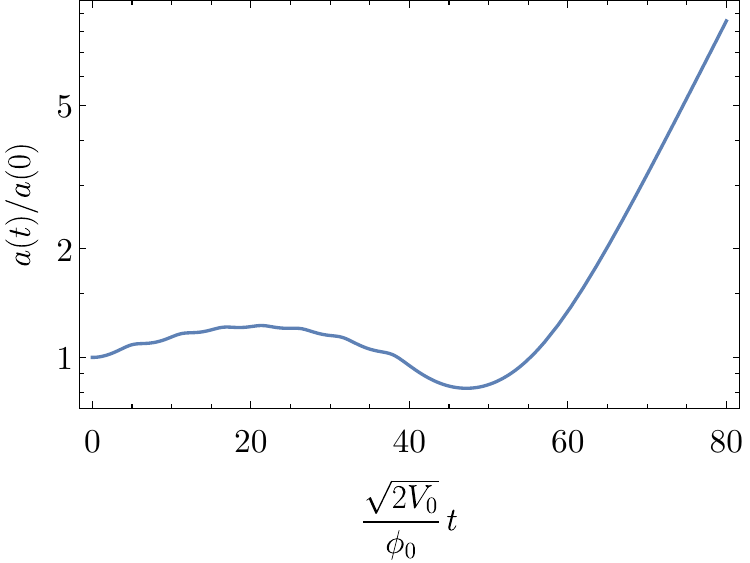}
    } \par
    \vskip 2mm
    \subcaptionbox{energy density components \label{sfig:dsp_rho}}{
    \includegraphics[width=0.7 \columnwidth]{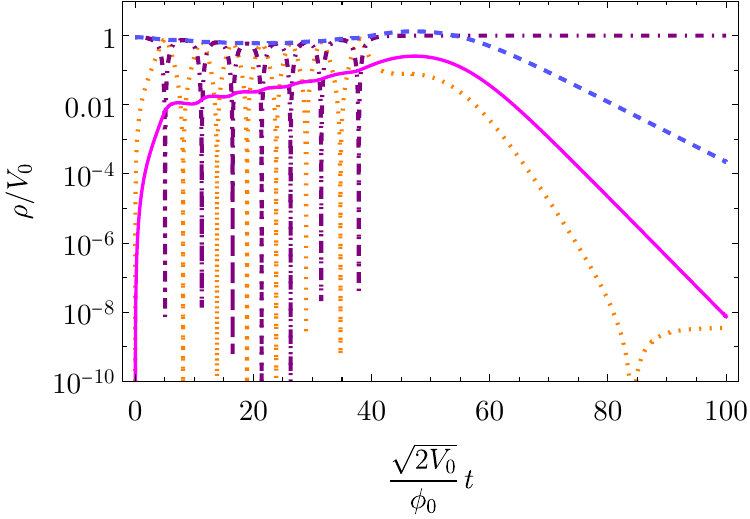}
    }
    \caption{A quasi-cyclic solution followed by inflation. \eqref{sfig:dsp_phi} The scalar field $\phi(t)$ in units of $\phi_0$. \eqref{sfig:dsp_a}  
    The logarithmic plot of the scale factor $a(t)$ in units of its initial value.
    \eqref{sfig:dsp_rho} Energy density components are plotted. The magenta solid line shows $\rho_\text{r}$, the purple dot-dashed line shows $V$, the dotted orange line shows the kinetic energy of $\phi$, and the light-blue dashed line shows $|\rho_K|$.   The parameters correspond to those of the bottom row of Fig.~\ref{fig:cyclic} ($\phi_0 = \sqrt{6} \times 10^{-1}$ and  $\phi_\text{ini}/\phi_0 = 1.8$), and additionally the dissipation rate is $\Gamma = 0.02 \sqrt{V_0}$.  The total e-folding number of inflation is larger than 300.  The effect of tachyonic instability is not included in this homogeneous simulation, but $\phi$ reaches the plateau $\phi/\phi_0 \gtrsim 3$ well before the fragmentation time observed in Fig.~\ref{fig:lattice}.  The radiation energy fraction becomes maximal $\rho_\text{r}/V_0 \approx 33 \%$ at around the bounce time whereas the gradient energy in Fig.~\ref{fig:lattice} is $\mathcal{O}(1\%)$, so the tachyonic instability will not qualitatively change the picture.}
    \label{fig:quasi-cyclic_dissipation}
\end{figure}
%%%%

Examples of the solutions are shown in Fig.~\ref{fig:quasi-cyclic_dissipation}.  The behavior of the solutions can be understood as follows. 
At the initial stage, the amount of radiation is negligible, and the cyclic solutions discussed in the previous section are good approximations.  When the fraction of $\rho_\text{r}$ becomes, say, $\mathcal{O}(10\%)$ of the total energy density, the effect of the radiation is no longer negligible. 
 The effect of radiation is more important around a bounce than around a turn-around into contraction because $\rho_\text{r} \propto a^{-4}$ ($w_\text{r} =1/3$) and the dynamics of $\phi$ is more susceptible around the plateau. Because of the blueshift of $\rho_\text{r}$, the timing of vanishing $H$ and hence the bounce are delayed in the contracting Universe. Consequently, $\rho_\phi$ also increases and the amplitude of $\phi$ can become significantly larger.  Once this happens, the slow-roll condition is well satisfied, and slow-roll inflation begins.  Note that the bounce requires accelerated expansion $\ddot{a}> 0$, so (long-lasting) inflation can naturally begin~\cite{Matsui:2019ygj}. After the onset of inflation, the spatial curvature and radiation are diluted by the quasi-exponential expansion, and the cyclic phase ends.
From the orange dotted line in Fig.~\ref{sfig:dsp_rho} showing the kinetic energy of $\phi$, one can see that the inflaton stops and changes the direction of motion at around $(\sqrt{2V_0}/\phi_0)\, t \approx 84$ and then approaches the slow-roll attractor solution. 

The above discussion also shows that when the growth of radiation (or any other positive energy component) is too fast, the bounce is further delayed. Since the kinetic energy of the scalar field scales like $\rho_\text{kin} \propto a^{-6}$ ($w_\text{kin} = 1$), it dominates the energy density. In this case, $\ddot{a}$ does not become positive and the universe collapses with the big crunch. To avoid it, $\Gamma$ should be sufficiently smaller than the typical scale of $H$, $H_\text{typical} \lesssim \mathcal{O}(H_\text{inf}) = \mathcal{O}(\sqrt{V_0/3})$, where $H_\text{inf}$ is the Hubble parameter during inflation. 

In this way, the inclusion of dissipative interactions can turn the cyclic solutions into solutions with a quasi-cyclic phase followed by the standard slow-roll inflation.  For one thing, the pure cyclic solutions (with small $a_\text{max}/a_\text{min}$) did not seem to have anything to do with the actual Universe, but the dissipation can turn these solutions very interesting and potentially relevant to our Universe.  For another, the tachyonic instability tends to spoil the cyclic Universe, but the dissipation can turn the fate of the Universe from the big crunch to inflation if the final bounce occurs before the fragmentation time.

\subsubsection{Where in the parameter space the mechanism works}

It is worthwhile to study where in our parameter space the above mechanism works.  We have four key parameters: the characteristic field value $\phi_0$, the overall factor or the height of the potential $V_0$, the initial scalar field value $\phi_\text{ini}$, and last but not least, the dissipation rate $\Gamma$.  In fact, these parameters control some important time scales, whose competition governs the dynamics of the whole system.  First, the time scale of an oscillation of $\phi$ is $\Thalf \sim \mathcal{O}(\phi_0 /\sqrt{V_0})$.  This also approximately gives the fragmentation time $t_\text{frag} \sim \mathcal{O}(\Thalf \ln (\phi_0^4 / V_0))$.   The time scale of an oscillation of $a$ is $\Thalfa \sim \Thalf / \phi_0 \sim \mathcal{O}(1/\sqrt{V_0})$.  More precisely, $\mathcal{O}(V_0)$ in these expressions are $V_0 - \rho$, which also depend on $\phi_\text{ini}$.  Finally, the time scale when the dissipation becomes relevant, i.e., when the quasi-cyclic solution is substantially modified  is controlled by $\Gamma^{-1}$.  Since exploring the four-dimensional parameter space is a bit complicated, let us proceed step by step. 

The first question is how generically the Universe, after being created from nothing, follows the (quasi-)cyclic behavior in the absence of dissipation. 
Later, we reintroduce dissipative effects.  Apart from the fragmentation time, $V_0$ just controls the overall time scale, so we can focus on the two-dimensional parameter space spanned by $\phi_0$ and $\phi_\text{ini}$.  The importance of the tachyonic instability is parametrized by the ratio $t_\text{frag}/\Thalf$, which depends only logarithmically on $V_0$. 

As the lattice computation for tachyonic instability is time-consuming, we study the homogeneous system and compare the obtained dynamics with the fragmentation time [Eq.~\eqref{fragmentation_time}] validated in the previous section. If some interesting dynamics occurs before the fragmentation time, it is interpreted to be valid with the inhomogeneous degrees of freedom.  On the other hand, if it occurs after the fragmentation time, it is regarded as an artifact of a homogeneous simulation. 

We have solved the equations of motion for $200 \times 200$ choices of the parameters in the range $-2 < \log_{10}\phi_0 \leq 0$ and $0 < \phi_\text{ini}/\phi_0 \leq 6$ with the time domain $0 \leq t \sqrt{V_0} \leq t_\text{max} \sqrt{V_0} = 300$. Solutions are automatically categorized into five classes as follows. 
\begin{itemize}
    \item When $\phi(t) \geq 5 \phi_\text{ini}$, $\dot{\phi} > 0$, and $\dot{a}(t) < 0$ are satisfied, the solution is classified as the big crunch with $\dot{\phi} > 0$.
    \item When $\phi(t) \leq -5 \phi_\text{ini}$, $\dot{\phi} < 0$, and $\dot{a}(t) < 0$ are satisfied, the solution is classified as the big crunch with $\dot{\phi} < 0$.
    \item If the scale factor remains in the range $a(0)/4 < a(t_\text{max}) < 4 a(0)$ at the end of the simulation $t=t_\text{max}$, the solution is classified as cyclic.
    \item If the e-folding number of expansion from the beginning of the simulation satisfies $\ln (a(t_\text{max})/a(0)) \geq 50$, the solution is classified as long enough inflation.
    \item When the solution is not classified into any of the above classes, they typically correspond to short inflation with $\ln (a(t_\text{max})/a(0)) < 50$. Therefore, we classify this class as short inflation.
\end{itemize}

%%%%%%%%%%%%%%%%%%
\begin{figure}[htb!] 
\begin{center}
\includegraphics[width=0.6 \columnwidth]{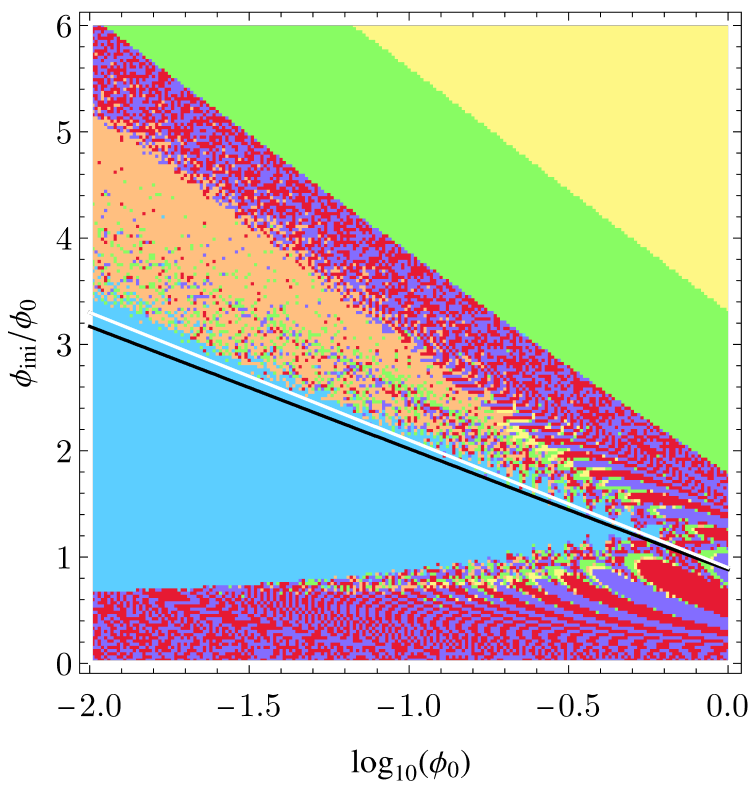}
\caption{Dependence of the type of solutions on the parameter $\phi_0$ and the initial condition $\phi_\text{ini}$ with dissipation turned off ($\Gamma=0$). The (1) \colorBigCrunchPositive, (2) \colorBigCrunchNegative, (3) \colorCyclic, (4) \colorShortInflation, (5) \colorLongInflationTachyonic{}, and (6) \colorLongInflation{} regions correspond to (1) big crunch with $\dot \phi > 0$, (2) big crunch with $\dot \phi < 0$, (3) cyclic, (4) short inflation, and long enough inflation with more than 50 e-foldings (5) after or (6) before the fragmentation time. We set $V_0$ to fit the CMB normalization. 
The boundary between the region (5) and (6) depends on $V_0$, but the other aspects do not depend on $V_0$. 
The parameter space on the \colorChoice{} line is further studied in Fig.~\ref{fig:generality} with the effects of dissipation.  
The first slow-roll parameter is $1$ on the \colorFlatness{} line, signaling the violation of the flatness condition below it. 
}
\label{fig:generality_basic}
\end{center}
\end{figure}
%%%%%%%%%%%%%%%%%%

Note that each category of the classification does not necessarily have a physically invariant notion. For example, most of the (short and long) inflationary solutions will end with a big crunch at (much) later time than $t_\text{max}$. In fact, the classification depends on $t_\text{max}$. Nevertheless, we find this analysis is useful to have at least a primary idea of what is going on at each parameter point. 

The result of the analysis is shown in Fig.~\ref{fig:generality_basic}. The \colorCyclic{} wedge-like region supports cyclic solutions. The region extends to the left of the figure, so cyclic solutions at the homogeneous level and in the classical framework are generic in our model.  As expected, there is also a long (\colorLongInflation) as well as a short (\colorShortInflation) inflation region around the top right corner.  In this part, $\phi_\text{ini}$ is sufficiently large so that inflation begins directly.  In our simulation, the cosmological constant is absent, so the inflated Universe will also eventually experience the big crunch due to spatial curvature.  When the e-folding number is sufficiently large, it does not occur within the simulation time. However, for a sufficiently short inflation case, the big crunch occurs within the simulation time.  In this way, the big crunch regions (\colorBigCrunchPositive{} for $\dot \phi > 0$ and \colorBigCrunchNegative{} for $\dot \phi < 0$) are adjacent to the short inflation region.  Interestingly, some curved stripe-like structures formed by the big crunch regions with $\dot{\phi} > 0$ and $\dot{\phi} < 0$ can be seen in the figure.  Somewhat surprisingly, between the cyclic and big crunch regions, we see some long (as well as short) inflation regions. A majority of such regions are panted by \colorLongInflationTachyonic{}, meaning that the starting time of inflation in the homogeneous simulation is later than the fragmentation time. In reality, these solutions represent big crunch solutions due to the fragmentation.  However, there are some \colorLongInflation{} regions.  In these regions, the Universe experiences bounce(s) and then inflation occurs. 
Thus, the quasi-cyclic Universe can turn into the inflationary Universe \emph{even without dissipation}. In these regions, however, the solution is susceptible to a very small perturbation, so some of them (close to the boundary of each colored region) may be affected by numerical errors. We leave a more dedicated study for future work. The existence of such regions suggests that the quasi-cyclic Universe may easily turn into the inflationary Universe with the help of small dissipation particularly for the phase boundary between the cyclic region and the adjacent long-inflation region.

Recalling the discussion around Eq.~\eqref{flatness_condition}, the region below the \colorFlatness{} line in Fig.~\ref{fig:generality_basic} where the slow-roll condition is violated,
\begin{align}
\frac{\phi_\text{ini}}{\phi_0} \lesssim \frac{1}{2} \log \left(\frac{4\sqrt{2}}{\phi_0}\right), 
\end{align}
corresponds to initial conditions of the classical Universe that are not favored by quantum cosmology analyses.  Nevertheless, we also show such a region to simply explore what kind of dynamics of the Universe occur from such classical initial conditions. 
Once we remove this part from our consideration, only the part close to the upper boundary (above the \colorFlatness{} line) of the \colorCyclic{} region can support cyclic solutions.

The second question is where in our parameter space the quasi-cyclic Universe turns into the inflationary Universe in the presence of dissipation.  
We consider the plane spanned by $\phi_0$ and $\Gamma$ (in units of $\sqrt{V_0}$) since these are the most important parameters.  $V_0$ can be fixed by fitting the CMB normalization $\phi_0 / \sqrt{V_0} = 2.1 \times 10^5$.  There is also $\phi_\text{ini}$ dependence. We are basically interested in $\phi_\text{ini}$ close to the boundary of the flat region.  In more detail, it should not be too small because (1) the flatness condition of quantum cosmology [cf.~Eq.~\eqref{flatness_condition}] would be violated, (2) $\sqrt{2V_0}/\phi_0 \times t_\text{frag}$ is larger for sufficiently large $\phi_\text{ini}/\phi_0$, and (3) it is expected that inflation easily occurs by small dissipation if the initial amplitude is larger.  Motivated by these facts, we choose a line close to the upper boundary between the cyclic  and non-cyclic regions in Fig.~\ref{fig:generality_basic}.  
Specifically, the following $\phi_\text{ini}$ is chosen as an example:
\begin{align}
    \frac{\phi_\text{ini}}{\phi_0} = 0.9 - 1.2 \log_{10} \phi_0. \label{phi_ini_choice}
\end{align}
This is shown by the \colorChoice{} line in Fig.~\ref{fig:generality_basic}. 
 We have also studied some other choices such as constant $\phi_\text{ini}/\phi_0$, on which we comment below.

%%%%%%%%%%%%%%%%%%
\begin{figure}[htbp!] 
\begin{center}
\includegraphics[width=0.6 \columnwidth]{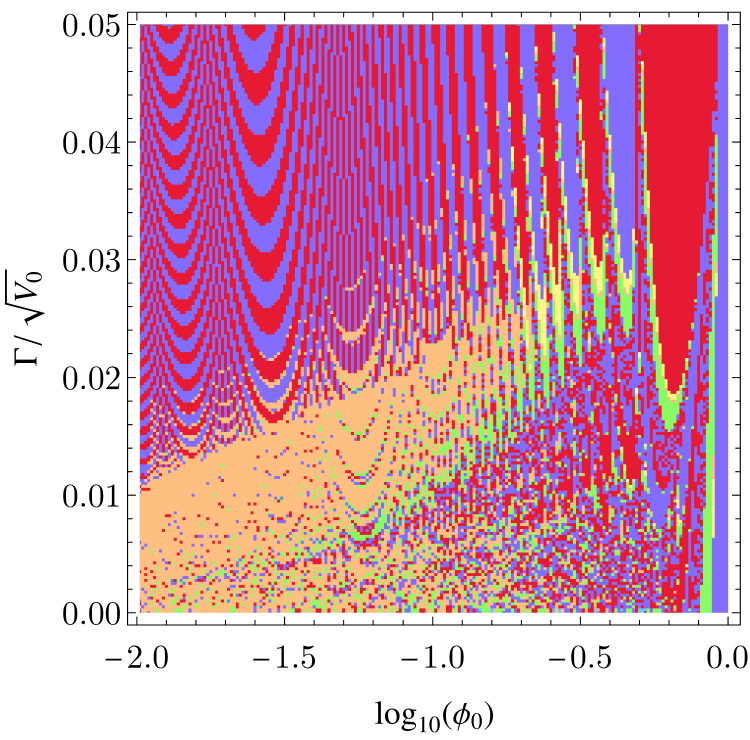}
\caption{Dependence of the type of solutions on the parameters $\phi_0$ and $\Gamma/\sqrt{V_0}$. $V_0$ is fixed to fit the CMB normalization, $\phi_0/\sqrt{V_0}=2.1 \times 10^5$. $\phi_\text{ini}$ is chosen to satisfy Eq.~\eqref{phi_ini_choice}.  The (1) \colorBigCrunchPositive, (2) \colorBigCrunchNegative, (3) \colorCyclic, (4) \colorShortInflation, (5) \colorLongInflationTachyonic, and (6) \colorLongInflation{} regions correspond to (1) big crunch with $\dot{\phi} > 0$, (2) big crunch with $\dot{\phi} < 0$, (3) cyclic universes, (4) short inflation, (5) long enough ($\log (a(t_\text{max})/a(0)) \geq 50$) inflation (neglecting tachyonic instability), and (6) long enough inflation before the fragmentation time, respectively. Note that some of the regions of type (5) become type (6) if we decrease $V_0$.}
\label{fig:generality}
\end{center}
\end{figure}
%%%%%%%%%%%%%%%%%%

We have performed an analysis similar to that for Fig.~\ref{fig:generality_basic}, in the presence of the constant dissipation $\Gamma$. The result of the analysis is shown in Fig.~\ref{fig:generality}. The \colorBigCrunchPositive{} and \colorBigCrunchNegative{} regions correspond to big crunch with $\dot{\phi}>0$ and $\dot{\phi} < 0$, respectively. The striped feature is clearer than the analogous feature in Fig.~\ref{fig:generality_basic}. This shows that when (at which phase of the oscillations of $\phi$) the kinetic-energy domination occurs depends significantly on $\phi_0$ and $\Gamma$. The \colorShortInflation{} region supports short inflation. It is intriguing that short inflation regions are narrower than in Fig.~\ref{fig:generality_basic} although it is not straightforward to compare different sections of the parameter space. Both the \colorLongInflationTachyonic{} and \colorLongInflation{} regions correspond to long enough inflation.  However, in the former regions, the time of the would-be-beginning of inflation due to the dissipation effect is after fragmentation due to tachyonic instability for the value of $V_0$ fitting the CMB normalization. This means that inflation does not occur in the \colorLongInflationTachyonic{} regions.  On the other hand, inflation successfully occurs before the fragmentation time in the \colorLongInflation{} regions. 

There are some additional comments on the figure.  First, note that the boundary between 
the \colorLongInflationTachyonic{} and \colorLongInflation{} regions moves to the left, or in other words, the \colorLongInflationTachyonic{} regions become \colorLongInflation{} and hence viable as $V_0$ decreases so that the fragmentation time increases compared to the time scale of the oscillations of $\phi$. To take this option, one has to abandon the CMB normalization by our potential and assume that the observable part of inflation is explained by other parts of the scalar potential. For example, the inflation responsible for the CMB may have occurred after the inflation we are discussing here. 

Second, we have studied other choices of $\phi_\text{ini}/\phi_0$ such as a constant value. In such cases, the details of the appearance of Fig.~\ref{fig:generality} change. For example, we see that the band of long-inflation regions has a small positive tilt, but this can become negative for other choices of $\phi_\text{ini}/\phi_0$. However, the rough qualitative picture is the same.  

Finally, we comment on a series of long-inflation regions.  In addition to the bulk regions, there are a series of disjoint long-inflation regions. Some \colorLongInflationTachyonic{} and \colorLongInflation{} arcs align with the boundary of the  Big-Crunch regions with $\dot{\phi} > 0$ and $\dot{\phi}<0$. This is because the kinetic energy of $\phi$ at the relevant time is suppressed at the boundary of the parameter space, and long inflation can occur more easily. On this ground and because of the continuity of the velocity of $\phi$ in terms of the underlying parameters, we expect there are a series of fine-tuned solutions for long-enough inflation even in the regions where we do not see them in Fig.~\ref{fig:generality} because of the resolution of the figure.

\subsubsection{Model building possibilities}

Let us discuss how we can realize the above setup in a more concrete quantum-field-theoretic framework. Although the dissipation rate in such a framework may depend on the field value $\phi$ and/or the temperature $T$, the constant $\Gamma$ case studied above is useful for order-of-magnitude estimates. First, we need to notice that the required size of the dissipation rate, $\Gamma \sim 0.02 \sqrt{V_0} \sim 10^{-2} m_\phi (\phi_0 / (\sqrt{6}\times 10^{-1}))$, where $m_\phi^2 =2V_0 / \phi_0^2$ is the inflaton mass at the minimum of the potential, is quite large to modify the quasi-cyclic solution before the fragmentation time.  If we consider simple perturbative decay of $\phi$, e.g., into fermions via a Yukawa coupling, it must be roughly of $\mathcal{O}(1)$, and the radiative corrections to the inflaton potential would be sizable.  

This motivates us to consider a shift-symmetric coupling that does not ruin the flatness of the inflaton potential. The approximate shift symmetry of the canonically normalized inflaton $\phi$ may have an origin such as scale invariance or U(1) symmetry. In the former case, we can naturally obtain the plateau potential as in $R^2$ inflation and Higgs inflation, and the dilaton-like coupling of the form $m_\psi e^{-\phi / f} \bar{\psi}\psi$ does not introduce dangerous quantum corrections~\cite{Bezrukov:2010jz, Kallosh:2016gqp}. In the latter case, the inflaton $\phi$ is axion-like and its potential may have a plateau region if the period (decay constant) is sufficiently large~\cite{Freese:1990rb,Czerny:2014wza,Higaki:2015kta}.  Alternatively, an infinite plateau as in our potential is possible also for an axion(-like particle) in a single branch of non-perturbative sectors: it is argued in Ref.~\cite{Nomura:2017ehb} that a plateau-type potential is generated if the axionic inflaton is coupled to the pure Yang-Mills theory. They proposed the inverse-hilltop potentials assuming a simple power law, but our potential is equally qualified as the axion potential in a single branch by dropping the power-law assumption. 

In both the dilatonic and axionic cases, the decay constant $f$ enters the decay rate as $\Gamma \sim m_\phi^3 / f^2$ or $m_\psi^2 m_\phi / f^2$, implying that $f$ needs to be small enough $f \sim \mathcal{O}(m_\phi)$.  
This is, however, in tension with a naive identification of the width of the potential $\phi_0$ with the decay constant, $\phi_0 \sim f$,\footnote{
In the model of Ref.~\cite{Nomura:2017ehb}, the relation between $\phi_0$ and $f$ is expected to be $\phi_0 \simeq 2 \pi f / \alpha'$ where $\alpha'$ is the fine structure constant of the gauge fields that generate $V(\phi)$.  These gauge fields should not be confused with the gauge fields we introduce in Sec.~\ref{ssec:model}.
} 
since we need $m_\phi \ll \phi_0$ to fit the CMB observables. 
In principle, the effectively small decay constant can be realized, $f = \phi_0 / c_a$, with a large anomaly coefficient $c_a \gg 1$~\cite{Anber:2009ua}.  
If $f$ is identified with the fundamental decay constant, on the other hand, the effective description will be questioned  when $H \gg f$ or $T \gg f$~\cite{DeRocco:2021rzv}. One or both of the inequalities are realized in the following discussion, which puts some constraints on the ultraviolet completion of the model. 

Another way to enhance the dissipation rate is to increase the number of decay channels.  This can be achieved by introducing various light fields and shift-symmetric couplings between $\phi$ and these fields. With $N$ such dissipation channels, the rate effectively scales as $N/f^2$, so the effective decay constant $f/\sqrt{N}$ can be lowered.  

Motivated by these discussions, we consider an axion-like inflaton $\phi$ and its coupling with non-Abelian gauge fields with the decay constant $f$ being a free parameter as a concrete example.\footnote{
This does not mean other model-building possibilities are excluded.  For example, we may consider coupling with an Abelian gauge field.  If the Abelian gauge field couples to charged particles, the potentially dangerous tachyonic instability of the gauge field (see Ref.~\cite{Adshead:2015pva} and references therein; Ref.~\cite{Garcia-Bellido:2016dkw} discusses some loopholes.) is suppressed~\cite{Domcke:2018eki}. 
}

%%%%%%%%%%%%%%%%%%%%%%%%%%%%%%%%%%%%%%%%
\subsection{Explicit model \label{ssec:model}}
We consider an axionic coupling between $\phi$ and non-Abelian gauge fields, 
\begin{align}
    \mathcal{L}_\text{int} =& - \frac{\alpha}{4\pi f} \phi \tilde F^{\mu\nu}F_{\mu\nu} , \\
    \mathcal{L}_\text{matter} =& - \frac{1}{4} g^{\mu\rho}g^{\nu \sigma} F_{\mu\nu} F_{\rho\sigma}, 
\end{align}
where $\tilde{F}^{\mu\nu} = \frac{1}{ 2 \sqrt{-g}}\epsilon^{\mu\nu\rho\sigma} F_{\rho\sigma}$ 
 is the Hodge dual of the field strength $F_{\mu\nu}$, $\alpha = g^2/(4\pi)$ is the ``fine structure constant'' for the gauge interaction with its coupling constant $g$, and $f$ is the effective axion decay constant.  For concreteness, we consider the gauge group SU($\Nc$).
In the numerical example below, we take $\Nc = 3$.

\subsubsection{Non-thermal dissipation}
The inflaton $\phi$ decays perturbatively into a pair of gauge bosons with a rate 
\begin{equation}
    \Gamma_\text{dec}= \frac{(\Nc^2-1) \alpha^2 m_\phi^3}{64 \pi^3 f^2}.
    \label{eq:decay}
\end{equation}
Precisely speaking, this perturbative decay rate is valid around the minimum of the potential.  A precise estimation of the gauge field production requires a preheating-like analysis, but the order of magnitude will be the same. 
With this in mind, we deal $m_\phi$ as the field-dependent mass ($m_\phi^2 = V''(\phi)$) in our numerical simulation.  Note that, when $V'' \leq 0$, $\Gamma$ as the decay rate of a particle with its mass $m_\phi$ does not make sense, and we turn off the decay channel.  In particular, the decay does not occur when $\phi$ is on a plateau. 
It is important to note that the inflaton energy dissipates through the perturbative decay over a significantly longer time scale than the oscillation period. This is due to $f > m_\phi$ and numerical suppression factors in $\Gamma_\text{dec}$. As previously discussed, the dissipation process must be effective before the onset of tachyonic instabilities, which occurs after several oscillation periods. To avoid the big crunch triggered by the tachyonic instabilities, therefore, a more efficient dissipation process is required.

\subsubsection{Thermal dissipation}
\label{sec:thermal-dissipation}
We generalize here our analysis by assuming the existence of a thermal bath consisting of a pure non-Abelian gauge field whose energy makes up a very small component of the total energy of the Universe when the Universe emerges from nothing. The thermal bath is characterized by temperature $T$ and we require that its energy is small enough that it plays no role in the dynamics of the background or the equation of motion of the inflaton for its first few oscillations 
around the bottom of its potential. However, we also assume that its energy is sufficiently large for the thermalization condition to be valid, namely the scattering rate of gauge bosons being sufficiently large $\Delta \equiv c_\text{thermal} \Nc^2\alpha^2 T \gg |H|$ where the numerical coefficient is assumed to be $c_\text{thermal} \simeq 10$ following Ref.~\cite{Laine:2021ego}.  We devote the next subsection to justifying  generation of the thermal bath in the immediate moments after the creation of the Universe. Such a thermal bath would give rise to additional thermal contributions to the dissipation rate, populating the Universe with radiation before the transition to inflation. These thermal effects are generally expected to dominate over the dissipation contribution due to decay in the case where $T\gg m_\phi$, which is the case we study in this section. It is possible that by considering thermal effects, one can achieve the transition to inflation for values of the inflaton-gauge field coupling that requires less hierarchy between the decay constant and the effective coupling to the gauge field. 

Such a scenario would give rise to dissipation due to rapid sphaleron transitions in the hot non-Abelian thermal bath. This sphaleron induced friction is well described by \cite{McLerran:1990de,Moore:2010jd,Bodeker:1999gx,Arnold:1999ux,Arnold:1999uy,Moore:2000mx,Moore:2000ara,Berghaus:2019whh,Berghaus:2020ekh,DeRocco:2021rzv}\footnote{
This formula is valid in the adiabatic regime. In our case, the adiabaticity condition $|\ddot{\phi}/\dot{\phi}| \ll \Delta$ (see, e.g., Ref.~\cite{Berghaus:2020ekh}) is marginally violated in the relevant time domain when $\Upsilon$ plays an important role, except for the boundaries of oscillations of $\phi$ when $\dot{\phi}$ vanishes and the left-hand side diverges instantaneously.  In such a case, the dissipation term in the equation of motion at time $t$ depends on the past dynamics at $t' < t$, or in other words, the linear term $\Upsilon \dot{\phi}$ needs to be augmented by higher time-derivative terms.  We leave the inclusion of such terms for future work.
}
\begin{equation}
    \Upsilon(T) \sim \frac{(\Nc \alpha)^5 T^3}{f^2}. 
    \label{eq:sphaleron}
\end{equation}
Additionally, one expects a contribution due to scattering between non-Abelian gauge quanta and inflaton quanta $A\phi \longleftrightarrow A A$~\cite{Takahashi:2019qmh, Moroi:2014mqa},
\begin{equation}
    \Gamma_\text{sct}(T) \simeq   \frac{C (\Nc^2-1)\alpha^2  T (p^0)^2}{4\pi^2 f^2 g^4} = \frac{C (\Nc^2-1) T (p^0)^2}{64\pi^4 f^2},
    \label{eq:scattering}
\end{equation}
where $C$ is an $\mathcal{O}(1)$ factor representing an uncertainty~\cite{Moroi:2014mqa}, which we take to be unity, and $p^0$ is the energy of a $\phi$ particle.\footnote{
The physical situation is different from that in Ref.~\cite{Moroi:2014mqa}, where the saxion dominating the energy density decays into axions, and the authors focused on the case in which the axion momentum is not too large $p \ll g^2 T$. In our case, $p^0$ is $\mathcal{O}(m_\phi)$ and $g^2 T$ rapidly increases above it. 
}   
We evaluate $p^0$ with a field-dependent mass of $\phi$ when it is positive.  When it is negative, it is unclear if this expression holds true, so we neglect the scattering rate for a conservative estimate of the thermal dissipation effects.  
These two contributions as well as the contribution due to decay given in Eq.~(\ref{eq:decay}) give the full dissipation rate defined in Eq.~(\ref{eq:scalareom}) in the toy model under consideration
\begin{align}
    \Gamma(T) = \Upsilon(T) + \Gamma_\text{sct}(T) + \Gamma_\text{dec}. \label{eq:total_decay_rate}
\end{align}

We have solved the set of equations \eqref{eq:scalareom}--\eqref{eq:friedman} with initial conditions (\ref{initial_condition}) supplemented with a small initial energy of the gauge field that is several orders of magnitude smaller than the initial potential energy of the axion. The results do not depend crucially on the initial energy of the radiation. For our parameters we choose the same benchmark parameters as in the examples presented in Sec.~\ref{ssec:dissipation_general}, namely $V_0 = (2.6 \times 10^{15}\,\text{GeV})^4$, $\phi_0 = \sqrt{6} \times 10^{-1}$, and $\phi_\text{ini}/\phi_0 = 1.8$.

\begin{figure}[htbp]
    \centering
    \subcaptionbox{field excursion \label{sfig:thermal_dsp_phi}}{
    \includegraphics[width=0.48\columnwidth]{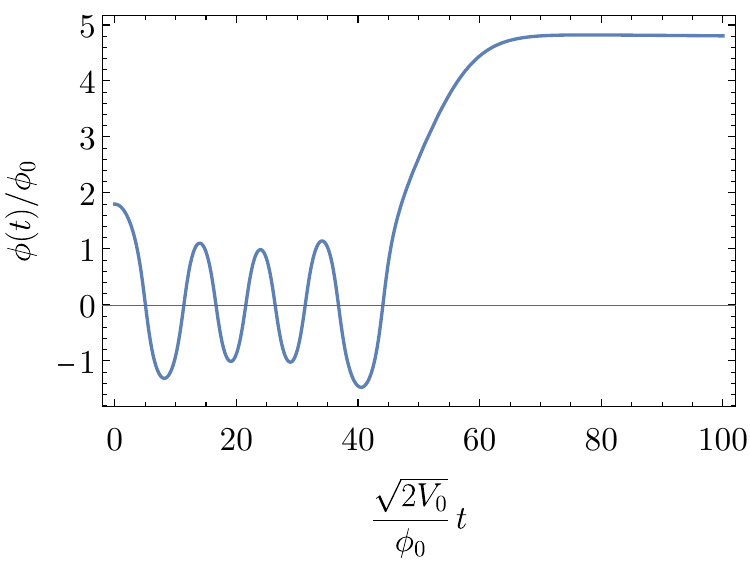}
    }~\subcaptionbox{energy density components \label{sfig:thermal_dsp_rho}}{
    \includegraphics[width=0.48\columnwidth]{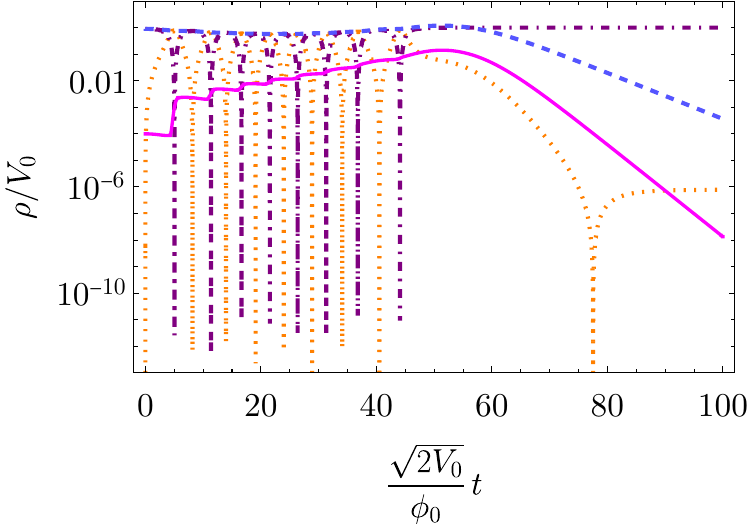}
    }
    \vskip 2mm
    \subcaptionbox{friction comparison \label{sfig:thermal_dsp_friction}}{
    \includegraphics[width=0.48\columnwidth]{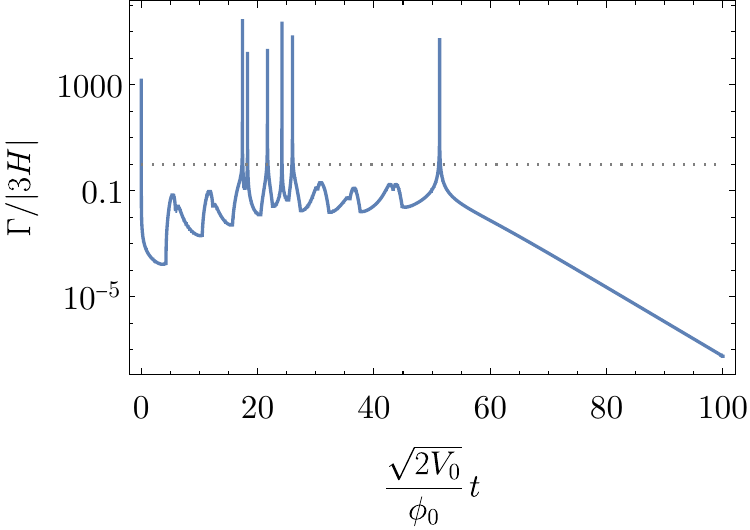}
    }~\subcaptionbox{thermalization condition \label{sfig:thermal_dsp_cond}}{
    \includegraphics[width=0.48\columnwidth]{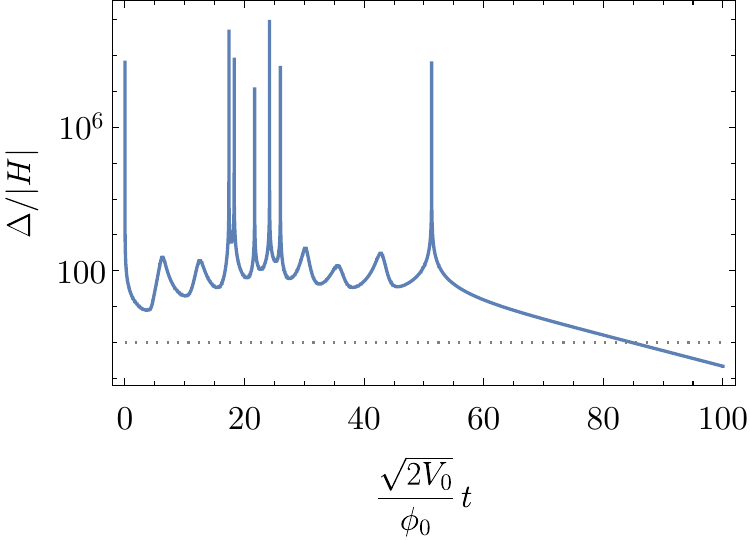}
    }
    \caption{A quasi-cyclic solution followed by inflation for the thermal case. \eqref{sfig:thermal_dsp_phi} The scalar field $\phi(t)$ in units of $\phi_0$. \eqref{sfig:thermal_dsp_rho} Energy density components are plotted. The magenta solid line shows $\rho_\text{r}$, the purple dot-dashed line shows $V$, the dotted orange line shows the kinetic energy of $\phi$, and the light-blue dashed line shows $|\rho_K|$. \eqref{sfig:thermal_dsp_friction} It is demonstrated that the backreaction to the scalar field equation of motion due to thermal friction is mostly negligible during the fast oscillation phase except for the moments in which $H\sim 0$ and the moment at which the system transitions to inflation.  \eqref{sfig:thermal_dsp_cond} The condition for thermalization $\Delta/|H|\gg 1$ is satisfied during the transition and hence our effective description is valid. 
     The parameters correspond to those of the bottom row of Fig.~\ref{fig:cyclic} ($\phi_0 = \sqrt{6}\times 10^{-1}$ and  $\phi_\text{ini}/\phi_0 = 1.8$), and additionally the axion-gauge coupling corresponds to $f = 2.4 \times 10^{-5} M_\text{P}$.  
     We are using as an example a pure non-Abelian SU(3) gauge field with a scalar field in the fundamental representation (see Sec.~\ref{sec:reheating}) so that $g_* = 22$.}
    \label{fig:thermal_quasi-cyclic_dissipation}
\end{figure}

Our strategy consists of fixing the gauge coupling at some representative value in the weak regime which we choose to be $\alpha=0.015$\footnote{
With this choice, the contribution to the potential of $\phi$ non-perturbatively generated from the coupling with the non-Abelian gauge fields is suppressed at least by a factor $\exp( - 2 / \alpha)\approx 10^{-58}$, so we neglect it.
} and then slowly increasing the value of the effective inflaton-gauge field coupling $f^{-1}$ until we achieve the transition from an oscillating Universe to an inflating one within at least $(\sqrt{2V_0}/\phi_0) \, t < 48$, which is the approximate time at which the gradient energy of the axion can no longer be neglected. After applying this procedure, we observe that we mostly obtain solutions for which the universe collapses and the kinetic energy of the scalar field increases indefinitely in one of the two directions. At the boundary of such solutions, we find solutions that gracefully transition to inflation. Obtaining such solutions within the time allowed is only possible with some moderate fine-tuning of the inflaton-gauge field coupling, and we present one such example in Fig.~\ref{fig:thermal_quasi-cyclic_dissipation}.

\subsubsection{Warming up the cold Universe}

The setup outlined in the previous subsection assumes that the non-Abelian gauge fields form a thermal bath quickly after the emergence of the Universe. Although a detailed study of thermalization is beyond the scope of this work, we nevertheless devote this subsection to demonstrate that our toy model contains all the necessary ingredients for the efficient formation of such a thermal bath due to the strong tachyonic instabilities present in the gauge field equations of motion, combined with the efficient self-interactions of the non-Abelian gauge field. We follow the set of criteria outlined in Ref.~\cite{DeRocco:2021rzv} and derive approximate analytic formulas that quantify whether the assumption of thermalization is valid during the early oscillation period, during inflation, and during the reheating era, respectively.

We decompose the gauge field operator in the Lorentz gauge as follows
\begin{equation}
    \bm{A}=\sum_{\lambda=\pm}\int \frac{\mathrm{d}^3 k}{(2\pi)^3}\left[A_\lambda (\tau,\bm{k})\bm{\epsilon}_\lambda (\bm{k} )a_\lambda(\bm k) \,{\rm e}^{i \bm{k}\cdot \bm{x} }+ {\rm H.c.}\right]\;, 
\end{equation}
where a bold-face letter denotes a three-dimensional vector, and the representation indices for the gauge algebra are omitted. $k$ is the magnitude of the wave-vector $k\equiv |\bm{k}|$ and the polarization vectors of massless spin-one fields satisfy $\bm{k}\cdot \bm{\epsilon}_\lambda(\bm{k})=0, \; \bm{k}\times \bm{\epsilon}_\lambda (\bm{k})=-i \lambda k \bm{\epsilon}_\lambda (\bm{k}),\; \bm{\epsilon}_\lambda(-\bm{k})=\bm{\epsilon}_\lambda(\bm{k})^*,\; \bm{\epsilon}_\lambda(\bm{k})^*\cdot\bm{\epsilon}_{\lambda^\prime}(\bm{k})=\delta_{\lambda,\lambda^\prime}$. 
Precisely speaking, expansion in terms of the spherical harmonics is more suitable than that in terms of the plane waves, but we omit such subtleties since the peak mode of the tachyonic instability of the gauge field is much larger than the fundamental infrared cutoff due to the spherical topology (see Appendix~\ref{sec:lattice}).
Once the Universe emerges from nothing, the gauge modes are at their vacuum configuration, and their evolution is well described by the equation of motion
\begin{equation}
    \ddot{A}_\pm+\left(k^2 \mp  \frac{k\, \alpha\, \dot{\phi}}{\pi f}\right)A_{\pm}=0\;.
\end{equation}
Because of the initially low occupation numbers, we neglect the nonlinear terms, which we expect will be subdominant, and also assume that the backreaction to the background is negligible. Our analysis here is based on the flat spacetime and we will compare the relevant time scales with the Hubble expansion or contraction rate below. 

The coupling of the gauge field to the axion-like inflaton modifies the dispersion relation of the circular modes, leading to one of the two polarizations (for fixed small values of $k$) becoming tachyonic and growing exponentially rapidly. Eventually, provided the tachyonic modes grow sufficiently fast and that strong backreaction does not quench the instability prematurely, the nonlinear terms may become important, which would lead to efficient energy exchange among the modes. One expects that the final configuration for the momentum distribution of the modes to be the thermal one as that is the one that maximizes the entropy of the system. 

Concretely, to form the thermal bath, we require the following~\cite{DeRocco:2021rzv}:
\begin{itemize}
    \item The gauge field value at which the nonlinearities become important $A_{\rm NL}$ is reached within a Hubble time $t\sim 1/H$.
    \item Backreaction to the background dynamics is negligible until the nonlinearity is achieved.
    \item Once the nonlinearities become important, the total energy stored in the gauge field is sufficient to account for a thermalization rate that is greater than the Hubble rate $\Delta>H$.
\end{itemize}

For simplicity, we assume that the inflaton field velocity $\dot{\phi}$ is constant and in practice, we will consider its slow-roll or root-mean-square value depending on whether we study inflation or the rapidly oscillating phases. The instability leads to an exponential growth of the mode function $A_\pm \simeq \frac{{\rm e}^{\beta t}}{\sqrt{2 k}}$ with the growth exponent $\beta=\sqrt{k(k_{\rm max}-k)}$ where the instability threshold $k_{\rm max}$  is equal to
\begin{equation}
    k_{\rm max}\equiv\frac{\alpha |\dot{\phi}|}{\pi f}\;.
\end{equation}

Expanding it around the momentum of maximum instability $k\sim k_{\rm max}/2$ using the saddle-point approximation $\beta=\frac{k_{\rm max}}{2}- \frac{\left(k-k_{\rm max}/2\right)^2}{k_{\rm max}}+{\cal O}\left[\left(k-k_{\rm max}/2\right)^4\right]$, we obtain
\begin{equation}
    A_\pm \sim \sqrt{\int \frac{\mathrm{d}^3 k}{(2\pi)^3}\frac{{\rm e}^{2\beta t}}{2k}}\simeq \frac{k_{\rm max}}{4\pi}\left(\frac{2\pi}{k_{\rm max} t}\right)^{1/4}{\rm e}^{k_{\rm max}t/2}\;.
\end{equation}
Additionally, Ref.~\cite{DeRocco:2021rzv} estimated the gauge field value at which nonlinearities become manifest as\footnote{A better estimation for the onset of nonlinearities can be made by comparing the loop contribution to the two-point function of the gauge field compared to the tree level
\begin{eqnarray}
    {\cal R}_A=\frac{\left\langle A A\right\rangle_{\rm loop}}{\left\langle A A\right\rangle_{\rm tree}}=\frac{\includegraphics[scale=0.15]{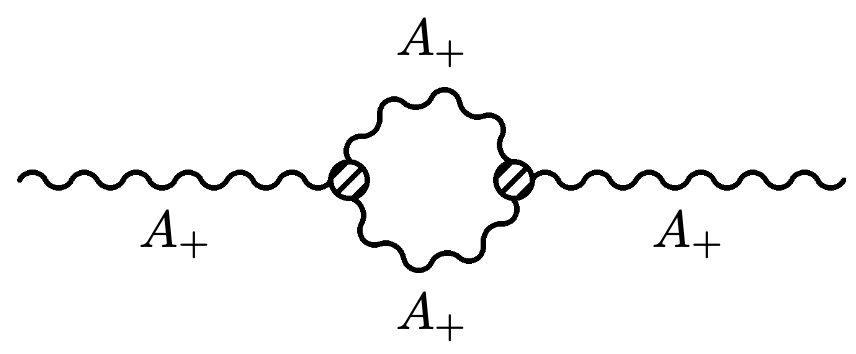}}{\includegraphics[scale=0.15]{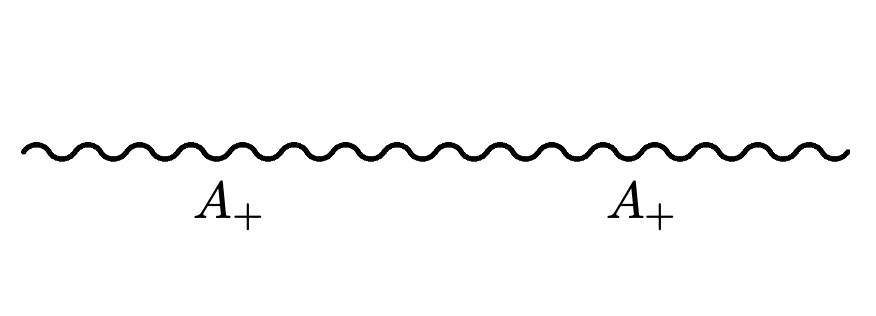}}
\end{eqnarray}
using the in-in formalism and finding the moment in time at which this ratio is equal to $1$ \cite{Peloso:2016gqs}. Because of the technical complications involved in computing the loop diagram, we leave this for future work and instead make an order of magnitude estimate as in Ref.~\cite{DeRocco:2021rzv}.
}
\begin{equation}
    A_{\rm NL}\sim \frac{k_{\rm max}}{2 \Nc g}\;.
\end{equation}

In order to facilitate our analysis, we also need approximate expressions for the energy density of the gauge field as well as for the backreaction term in the background equation of motion of the axion. Using again the saddle-point approximation, we obtain
\begin{equation}
    \rho_r\equiv\left\langle \frac{\bm{E}^2+\bm{B}^2}{2}\right\rangle\sim (\Nc^2-1)\int \frac{\mathrm{d}^3 k}{(2\pi)^3}\frac{k\,k_{\rm max}}{2}\frac{{\rm e}^{2\beta t}}{2 k}\simeq \frac{(\Nc^2-1)k_{\rm max}^2 A^2}{4}
\end{equation}
and
\begin{equation}
    {\cal B}\equiv\frac{\alpha}{\pi f}\left\langle \bm{E} \cdot \bm{B}\right\rangle\sim \frac{\alpha}{\pi f}(N_c^2-1)\int \frac{\mathrm{d}^3 k}{(2\pi)^3}\beta k\frac{{\rm e}^{2\beta t}}{2 k}\simeq \frac{\alpha (N_c^2-1)}{4\pi f}k_{\rm max}^2 A^2
\end{equation}
respectively. The ``electric'' and ``magnetic'' fields $\bm E$ and $\bm B$ are defined similarly to the usual Abelian electromagnetic fields.  Equipped with the expressions above, we may proceed to evaluate the conditions individually.

\paragraph{Condition 1:} The requirement that nonlinearities emerge within a Hubble time is rewritten as follows
\begin{equation}
 t_{\rm NL}< t_H
\end{equation}
or equivalently
\begin{equation}
    k_{\rm max}\equiv\frac{\alpha \vert\dot{\phi}\vert}{\pi f}\gg 2 H\log\left(\frac{2\pi}{\Nc g}\right)\;.
    \label{eq:cond1}
\end{equation}

\paragraph{Condition 2:} In order to ensure that the backreaction is negligible until at least the moment nonlinearities become important, Ref.~\cite{DeRocco:2021rzv} compared the energy of the gauge field to the kinetic energy of the inflaton. We show here that a stricter condition arises if we compare the backreaction term in the equation of motion of $\phi$ to the potential gradient instead. Note that the potential gradient is always parametrically the dominant term before, during, and after inflation.  Namely, we have the condition 
\begin{equation}
    {\cal B}(t_{\rm NL})\ll V'[\phi(t_{\rm NL})]
\end{equation}
or written in terms of $k_{\rm max}$
\begin{equation}
 k_{\rm max}\ll 2\left(\frac{4\pi^2 f \Nc^2 V'[\phi(t_{\rm NL})]}{(\Nc^2-1)}\right)^{1/4}.
 \label{eq:cond2b}
\end{equation}

\paragraph{Condition 3:}
Finally, the last requirement boils down to
\begin{equation}
    \Delta(t_{\rm NL})>H(t_{\rm NL})\;\;\;\;{\rm or}\;\;\;\;\rho_r\gg\frac{\pi^2 g_*}{30}\left(\frac{H}{\Delta}\right)^4,
\end{equation}
where $g_* = g_*(T)$ is the effective number of relativistic degrees of freedom, 
or in terms of the threshold momentum $k_{\rm max}$, 
\begin{equation}
    k_{\rm max}\gg \left(\frac{2\pi^3 g_*}{15 \alpha^3 \Nc^6(\Nc^2-1)}\right)^{1/4}\frac{H}{5}\;.
    \label{eq:cond3}
\end{equation}

The set of conditions (\ref{eq:cond1}), (\ref{eq:cond2b}), and (\ref{eq:cond3}) can be checked before, during, and after inflation through the replacements:
\begin{align}
    \dot{\phi}\rightarrow &\,  \frac{\phi_{0}}{\Thalf}  & {\rm and}&  & H\rightarrow& \, \frac{1}{\Thalf}  & &({\rm before\;inflation})\nonumber\\
    \dot{\phi}\rightarrow & \, \frac{\sqrt{V_0}\phi_0}{2\sqrt{3} M_\text{P} \Ne} & {\rm and}& &  H\rightarrow & \, \frac{\sqrt{V_0}}{\sqrt{3}M_\text{P}} & &({\rm during\;inflation})\nonumber\\
    \dot{\phi}\rightarrow & \,  m_\phi \phi_0  & {\rm and}& &  H\rightarrow & \, \frac{\sqrt{V_0}}{\sqrt{3}M_\text{P}}& &({\rm after\;inflation})
\end{align}
where $\Thalf$ in the first line is the half period of oscillation of $\phi$ defined in Appendix \ref{sec:osc_avg} and $\Ne$ in the second line is the number of e-folds left until the end of inflation.

All the conditions outlined in the previous paragraphs are easily satisfied for the set of parameters in our example in Sec.~\ref{sec:thermal-dissipation}. In fact, we find that thermalization occurs quickly within a time $t_{\rm NL}=0.06 \,\Thalf$ and therefore within about three percent of a single 
oscillation of $\phi$ around its minimum. At the time when nonlinearities become important, the backreaction term in the equation of motion of $\phi$ is an order of magnitude smaller, and the energy density of the gauge field is $\sim 10^{-4}$ part of the total energy of the Universe at that time with sufficiently large thermalization rate. Note that our finding for the energy density of the gauge field upon thermalization is fully consistent with the initial condition set during our numerical simulations in the previous section. We conclude that our assumption of the rapid formation of a thermal bath after inflation is valid and an inevitable consequence of the strong self-interactions of the non-Abelian gauge field combined with the strong tachyonic instabilities that form due to the coupling to the axion-like inflaton.

\subsubsection{Evolution during inflation}

After the transition to inflation, there are generally two distinct possibilities for the type of inflation one should expect. It is possible that the thermal bath and the inflaton  equilibrate in such a way that the thermalization condition is still satisfied, in which case the phenomenology of \textit{warm inflation}~\cite{Berera:1995ie} should apply. In the other case, the thermal bath decays away beyond the point at which the thermalization condition applies. The second scenario leads to \textit{cold inflation}. 

In our scenario, the thermal bath typically decays away as radiation after the onset of inflation. This decay, however, cannot continue indefinitely since there is always some radiation that is being produced due to the slow-roll motion of the inflaton. 
The steady-state equilibrium solution corresponds to a “radiation floor” which can be estimated by assuming slow-roll solutions and neglecting the highest derivatives in the evolution equations for the inflaton and radiation energy density respectively. Additionally, the only relevant source of friction during the inflationary phase is sphaleron-induced thermal friction, and we assume that inflation occurs at the “weak warm regime”~\cite{Graham:2009bf, Bastero-Gil:2011rva, DeRocco:2021rzv} in which the Hubble rate is much greater than the thermal friction, and therefore the dynamics of the inflaton background are unaltered due to the presence of the thermal bath. This is typically the case in our scenario since we only want the thermal friction to affect the motion of the axion momentarily just before the onset of inflation. The steady-state solution for the radiation temperature is then simply
\begin{eqnarray}
    T=\frac{5 \sqrt{3}\left(\Nc \alpha\right)^5\sqrt{V_0}\phi_0^2}{16 \pi^2  \left(\Nc^2-1\right)\Ne^2 f^2 M_\text{P}}.
\end{eqnarray}
This formula, in combination with the thermalization rate condition $\Delta\gg H$, can be used as a diagnostic to determine whether the thermal bath survives after the onset of inflation or whether it redshifts to a point where the thermalization rate is smaller than the Hubble rate, in which case we enter the cold inflation regime.

For the parameters of our example in Sec.~\ref{sec:thermal-dissipation}, the thermal bath indeed decays away to a point where the thermalization approximation ceases to be
 valid. We conclude that inflation occurs in the cold regime, at least in the early  stages of inflation. It is possible that in the last few e-folds of inflation, when the inflaton begins to speed up, the Universe may heat up again. This can be determined using the discussion in the previous section and indeed for our parameters inflation enters the weak warm regime in the last few e-folds before the end of inflation. This hardly has any phenomenological consequences since the scales that exit the horizon at that time are far smaller than CMB scales, for which we have significant constraints. 

%%%%%%%%%%%%%%%%%%%%%%%%%%%%%%%%%%%%%%%%%%%%%%%%%%%
\section{CMB observables and graceful exit from inflation \label{sec:CMB}}
We briefly mention the inflationary observables in our setup.
A priori, we do not need to identify the inflationary period immediately after the bounce and the inflationary period responsible for our CMB data.  However, it is interesting if we can identify them with each other as a minimal scenario. 
The potential~\eqref{potential} is identical to the simplest T-model of $\alpha$-attractor via the identification $\phi_0 \equiv \sqrt{6\alpha_\phi}$, and its prediction is~\cite{Kallosh:2013hoa, Ferrara:2013rsa, Kallosh:2013yoa, Galante:2014ifa, Carrasco:2015pla}
\begin{align}
    A_\text{s} = & \frac{\Ne^2 V_0}{3 \pi^2 \phi_0^2} \left(= \frac{\Ne^2 V_0}{18\pi^2 \alpha_\phi}\right), &
    n_\text{s} = & 1 - \frac{2}{\Ne}, &
    r =& \frac{2 \phi_0^2}{\Ne^2} \left( =  \frac{12 \alpha_\phi}{\Ne^2} \right), 
\end{align}
where $A_\text{s}$ and $n_\text{s}$ are the amplitude and the spectral index of the primordial curvature perturbations, $r$ is the tensor-to-scalar ratio, and $\Ne$ is the e-folding number between the horizon exit of the CMB scale and the end of inflation. As we have already emphasized, the choice of the potential in~\eqref{potential} is for demonstration, so other potentials and corresponding different inflationary observables are also possible.

As we have seen, once long-lasting inflation happens, the energy density and temperature of the gauge fields drop significantly. For sufficiently small values of $\Upsilon (\ll H)$, the temperature can drop below the Hubble rate, $T \ll H$, by the time the inflaton approaches the inflationary attractor solution, which corresponds to the value of the horizontal axis (dimensionless time) around 90 in Fig.~\ref{sfig:dsp_rho}. That is, the thermal environment is relevant only up to the transition regime, and the standard cold inflation is realized afterward.  
Soon after, the temperature  drops further, $\Delta \ll H$, and thermalization cannot be maintained, so the gauge fields will be diluted away to become negligible. We can use the above standard formulas for the inflationary observables.

The observational constraint $0.96 \lesssim n_\text{s} \lesssim 0.97$~\cite{Planck:2018jri} can be easily satisfied for the standard values of $\Ne$ ($50\lesssim \Ne \lesssim 60$).  The upper bound $r \lesssim 0.04$~\cite{BICEP:2021xfz, Tristram:2021tvh, Paoletti:2022anb} gives us a super-Planckian upper bound on $\phi_0$, which is automatically satisfied for the interesting parameter space $\phi_0 \lesssim 1$, where the (multi)bounces followed by inflation are possible. On the other hand, it is more nontrivial to fit the measured value of $A_\text{s}\simeq 2.1 \times 10^{-9}$~\cite{Planck:2018jri}. For $\phi_0 \ll 1$, the tachyonic instability becomes so strong and $V_0$ should be exponentially smaller to compensate for it, hence the $A_\text{s}$ value cannot be fit.  For $\phi_0 \gtrsim 1$, on the other hand, the bounce requires significant tuning.  Nevertheless, there is a window around $\phi_0 \sim 0.25$ in which observation-compatible inflation occurs due to dissipation barely before the fragmentation by the tachyonic instability would occur.  

In the simplest scenario, the inflaton field value goes back to the region around the minimum, where the (multi)bounces occurred before inflation.  Thus, the graceful exit from as well as graceful entry to inflation is realized in our setup. 

%%%%%%%%%%%%%%%%%%%%%%%%%%%%%%%%%%%%%%%%%%%%%%%%%%%
\section{Reheating of the Universe \label{sec:reheating}}

An interesting aspect of our scenario is that the reheating mechanism is predicted once we fix the mechanism of the graceful entry, and vice versa. If we put additional interactions for reheating, it would also affect the transition from the quasi-cyclic period to inflation.  

%%%%%%%%%%%%%%%%%%%%%%%%%%%%%%%%%%%%%%%%
\subsection{Preheating}

\begin{figure}[htbp] 
\begin{center}
\includegraphics[width=0.7 \columnwidth]{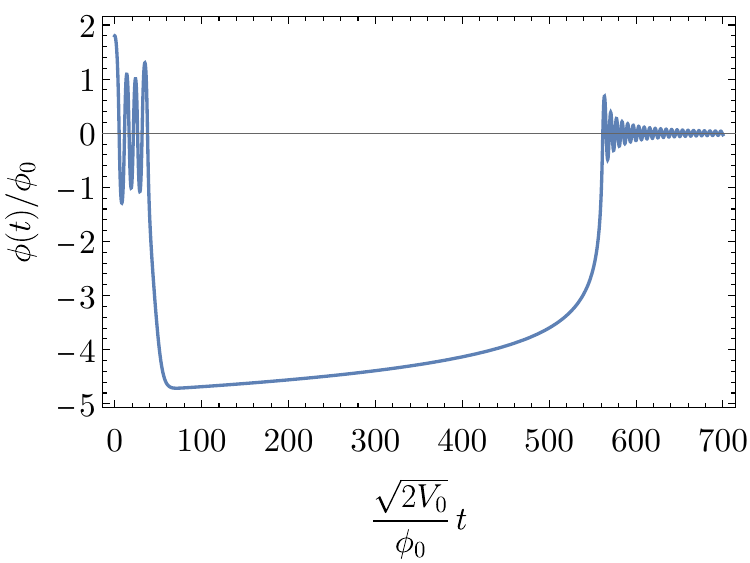}
\caption{
The solution for $\phi(t)$ throughout the bounce, inflation, and reheating epochs. For simplicity and to show the whole dynamics clearly, the toy model of a constant $\Gamma = 0.019 \sqrt{V_0}$ is used. In this case, more than $50$ $\mathrm{e}$-foldings of inflation occur.
}
\label{fig:long-time}
\end{center}
\end{figure}

As the tachyonic instability and the parametric resonance occur during the quasi-cyclic phase, one may expect that they occur again during the coherent oscillations of the inflaton $\phi$ after inflation.  Since the spatial curvature is essentially negligible after a long enough inflation, the physics is not the repetition of what occurred before inflation. The dynamics after inflation is the same as what is expected in the standard inflationary cosmology possibly with some thermal effects. Fig.~\ref{fig:long-time} shows it explicitly. 

%%%%%%%%%%%%%%%%%%
%%%%%%%%%%%%%%%%%%

Preheating after inflation~\cite{Dolgov:1989us, Traschen:1990sw, Shtanov:1994ce, Kofman:1997yn} with the plateau potential~\eqref{potential}  has been studied in Ref.~\cite{Tomberg:2021bll}.  Because of the Hubble friction after inflation, the field amplitude decreases significantly within a half oscillation of $\phi$ and the shape of the potential looks more and more purely quadratic, i.e., that of a free scalar field. 
We observe neither tachyonic instability nor  parametric resonance for $\phi_0 = \sqrt{6} \times 10^{-1}$ in simulations with \CosmoLattice. For these simulations, effects of dissipation of $\phi$ into other fields are not included to focus on the potential effects of tachyonic or parametric resonance.  This result is consistent with Ref.~\cite{Lozanov:2017hjm}.

For a concrete model, we have discussed $\phi$ interacting with non-Abelian gauge fields in Sec.~\ref{sec:dissipation}.  Gauge fields also develop tachyonic instability in the presence of the $\phi F \tilde{F}$ coupling and nonzero $\dot{\phi}$.  The energy of the inflaton is efficiently transferred to the non-Abelian gauge fields through the tachyonic instability.  
Since the interaction is strong enough, the non-linearity of the interactions becomes quickly important, and the system is expected to be thermalized. 
Once the gauge fields are thermalized, the inflaton energy density can dissipate through scattering in the thermal bath and the sphaleron effect on top of the usual decay effect as we discussed in Sec.~\ref{sec:dissipation}. 

%%%%%%%%%%%%%%%%%%%%%%%%%%%%%%%%%%%%%%%%
\subsection{Reheating}

\subsubsection{Reheating the dark sector}
The dissipation rate of the inflaton in the thermal background is given by Eq.~\eqref{eq:total_decay_rate}, i.e., $\Gamma(T) = \Upsilon + \Gamma_\text{sct} + \Gamma_\text{dec}$.  The reheating temperature of the non-Abelian gauge fields $T_\text{R}^\text{dark}$ can be estimated as in Ref.~\cite{Takahashi:2019qmh}.  In our case, $\Upsilon \ll H$ is always satisfied and the sphaleron transition process is less and less effective after inflation, so it is irrelevant for our purpose. The other two dissipation channels are more and more relevant as time goes on.  The reheating temperature of the dark sector $T_\text{R}^\text{dark}$ is  
\begin{align}
    T_\text{R}^\text{dark} = \max \left[ T_\text{R,\,dec}, \min \left[  T_\text{R,\,sct}, T_\text{max} \right] \right],
\end{align}
where $T_\text{R,\,sct}$ ($T_\text{R,\,dec}$) is the reheating temperature if the $\Gamma_\text{sct}$ ($\Gamma_\text{dec}$) balances with the Hubble rate $H$, and $T_\text{max}$ is the attainable maximal temperature after inflation, which is upper-bounded by the would-be reheating temperature in the instantanous reheating case, 
\begin{align}
T_\text{max} \leq T_\text{inst} \simeq & (30/(\pi^2 g_*) V_0)^{1/4} \nonumber \\
=& 1.6 \times 10^{15} \, \text{GeV}\left(\frac{g_{*}}{22}\right)^{-1/4} \left( \frac{V_0^{1/4}}{2.6 \times 10^{15} \, \text{GeV}} \right).
\end{align}
It is beyond the scope of our work to precisely estimate $T_\text{max}$. 
The expressions of $T_\text{R, sct (dec)}$ are given as follows
\begin{align}
    T_\text{R,\,sct} = & \left(\frac{90}{\pi^2 g_*}\right)^{1/2} \frac{C(\Nc^2-1) m_\phi^2 M_\text{P}}{64\pi^4 f^2} \nonumber  \\
    =& 1.5 \times 10^{14}\,\text{GeV} \left( \frac{C}{1} \right) \left(\frac{g_*}{22}\right)^{-1/2} \left(\frac{\Nc^2-1}{3^2-1}\right) \nonumber \\
    & \qquad \qquad \qquad \qquad \times \left(\frac{m_\phi}{1.6 \times 10^{13}\,\text{GeV}}\right)^2 \left(\frac{f}{5.8 \times 10^{13}\,\text{GeV}} \right)^{-2},\\
    T_\text{R,\,dec} =& \left( \frac{90}{\pi^2 g_*} \right)^{1/4} \frac{\sqrt{\Nc^2-1} \alpha m_\phi^{3/2} M_\text{P}^{1/2}}{8 \pi^{3/2} f} \nonumber \\
    =& 1.3 \times 10^{12} \, \text{GeV} \left(\frac{g_*}{22}\right)^{-1/4} \left(\frac{\Nc^2-1}{3^2-1} \right)^{1/2} \left( \frac{\alpha}{0.015} \right) \nonumber \\
    & \qquad \qquad \qquad \qquad \times \left( \frac{m_\phi}{1.6 \times 10^{13} \, \text{GeV}} \right)^{3/2} \left( \frac{f}{5.8 \times 10^{13}\,\text{GeV} }\right)^{-1} .
\end{align}
We see that $T_\text{R,\,sct}$ dominates over $T_\text{R,\,dec}$ for typical parameter values.

\subsubsection{Heating the Standard Model}

After reheating of the (dark) gauge fields, some interaction must be introduced to heat up the Standard Model (SM) sector, so the $T_\text{R}^\text{dark}$ roughly serves as an upper bound on the reheating temperature of the SM sector, 
    $T_\text{R} \lesssim T_\text{R}^\text{dark}$. 
It will be much smaller than this upper bound.  Non-Abelian gauge fields can only have a coupling with the SM sector through interaction terms of dimension 6 or higher, so they cannot be in equilibrium with the SM sector unless the cutoff scale of that interaction is sufficiently low. 

On the other hand, if the non-Abelian gauge interaction becomes strong and dark glueballs are formed, it is possible for the dark glueballs to decay into SM particles through this high-dimensional interaction. Considering the dimension 6 interaction between the SM Higgs and non-Abelian gauge fields, if the cutoff scale is around the Planck mass, the mass of the dark glueball must be heavier than $10^{10}$ GeV to reheat the SM sector before big bang nucleosynthesis. Another possibility is to introduce hidden matter with non-Abelian gauge interactions. Fermions have a significant impact on the thermal frictional force in our scenario~\cite{Berghaus:2020ekh}, so we consider introducing dark Higgs. In this case, a portal coupling between the dark Higgs and the SM Higgs can be introduced. As long as the coupling constant is not too small, it is expected that if the dark Higgs is thermalized, the SM sector will also be thermalized soon through this Higgs portal coupling. 

The specific process of heating the SM sector is model-dependent, but once the non-Abelian gauge sector is thermalized, it is not difficult to heat up the SM sector coupled to it. Also, in this scenario, the SM sector must always couple to the hidden gauge sector, so if the mass scale is sufficiently small, it is expected that various experimental and observational searches and constraints can be imposed, but this is beyond the scope of this paper.

%%%%%%%%%%%%%%%%%%%%%%%%%%%%%%%%%%%%%%%%%%%%%%%%%%%
\section{Wave function of the Universe and probabilistic implications \label{sec:creation}}

In this section, we discuss the initial conditions of our cosmological solutions in more detail. We saw that the presence of the tachyonic instability of $\phi$ precludes the cyclic period from being eternal to the future.  Similarly, it is not possible for the cyclic period to be eternal to the past since perturbations cannot decrease arbitrarily in quantum theory.  We have assumed that the initial value of $\phi$ is close to the edge of the plateau with sizable spatial curvature, so it is inconsistent to assume a long period of inflation occurred just before the quasi-periodic phase. Perhaps, the only natural way for the Universe to enter the quasi-periodic phase could be the creation of the Universe from nothing. In the following, we discuss how likely this initial condition should be by introducing the cosmological wave function $\Psi$, which represents the quantum nature of the spacetime and gives the nucleation probability of the Universe.

In the so-called minisuperspace approximation, 
we consider only the scale factor $a(t)$
and the homogeneous scalar field $\phi(t)$, and the cosmological wave function $\Psi(a,\phi)$
satisfies the Wheeler-DeWitt equation~\cite{DeWitt:1967yk}, 
\begin{equation}\label{WDW}
{\cal H}(a,\phi) \Psi(a,\phi) = 0 ,
\end{equation}
with the Hamiltonian operator ${\cal H}$ written as~\cite{Kiefer:2008sw}
\begin{equation}\label{Hamiltonian}
{\cal H}(a,\phi) = -\frac { 1 }{ 12 v_K \, { a } } \frac { \partial ^2 }{ \partial a^2 }  +\frac { 1 }{ 2 v_K \, { a }^{ 3 } } \frac { { \partial  }^{ 2 } }{ \partial { \phi  }^{ 2 } }+{ a }^3 v_K \left( \frac{3K}{a^2}-V\left( \phi  \right)  \right),
\end{equation}
where $v_K = 2 \pi^2 K^{-3/2}$ is the three-dimensional conformal volume of the spacelike hypersurface, and we neglect the subtlety of the operator ordering ambiguity of $a$~\cite{Hartle:1983ai} preferring a simple expression since it is irrelevant for the wave function in the semiclassical regime.

The potential barrier in the Wheeler-DeWitt equation suggests that
the Universe might have emerged from nothing ($a = 0$)
by quantum tunneling. The Universe created in this way 
should have a finite size and obey the initial 
conditions~\eqref{initial_condition}~\cite{Vilenkin:1987kf}. 
In the preceding sections, we have adopted Eq.~\eqref{initial_condition} as the initial condition of the subsequent classical time evolution.  How the quantum Universe dynamically becomes classical by using methods developed in quantum cosmology is discussed in Refs.~\cite{Hartle:2008ng,  Battarra:2014xoa, Battarra:2014kga, Lehners:2015sia, Lehners:2015efa}.

In Eq.~\eqref{Hamiltonian}, we have omitted terms corresponding to $\mathcal{L}_\text{int}$ and $\mathcal{L}_\text{matter}$ in Eq.~\eqref{full_theory}.  It is possible to include the effects of such terms.  For example, we may consider thermal quantum creation of the Universe by including the radiation energy density $\rho_\text{r} = \epsilon_\text{r} a^{-4}$ with $\epsilon_\text{r}=$ const., which might be quantized~\cite{Damour:2019iyi}.  In this case, the Wheeler-DeWitt equation for the wave function has the same form as the standard energy eigenvalue equation,
\begin{align}
\label{eqn:WDWeq}
& \left[-\frac { 1 }{ 12 } \frac { \partial^2  }{ \partial a^2 } +\frac { 1 }{ 2{ a }^{ 2 } } \frac { { \partial  }^{ 2 } }{ \partial { \phi  }^{ 2 } }+ { a }^4v_K^2\left( \frac{3K}{a^2}-V\left( \phi  \right)  \right) \right]\Psi(a,\phi) = \epsilon_\text{r} v_K^2 \Psi(a,\phi) \, .
\end{align}
A potentially interesting implication of such a preexisting radiation component for the subsequent classical cosmological dynamics is that the necessary magnitude of dissipation may be relaxed.  Quantitative analyses for this point are left for future work.

\subsection{Probabilistic implications}
The nucleation probability of the Universe depends on the boundary conditions, and there are several proposals for them in quantum cosmology: the tunneling proposal~\cite{Vilenkin:1984wp}, the DeWitt proposal~\cite{DeWitt:1967yk}, and the no-boundary proposal~\cite{Hartle:1983ai}.
Originally, the tunneling proposal was formulated in terms of the Wheeler-DeWitt equation, while the no-boundary proposal was formulated 
based on the Euclidean path integral.  
 However, both of them can be formulated in either formalism, and there is a long-lasting controversy about which proposals should be taken. 
 Following Refs.~\cite{Vilenkin:1986cy,Vachaspati:1988as}, 
we can estimate the tunneling or no-boundary
wave function using the Wentzel-Kramers-Brillouin (WKB) approximation,
\begin{align}
\Psi(a, \phi) \simeq e^{-S[a,\phi]},
\end{align}
where $S[a,\phi]$ is the WKB action at the leading order written as 
\begin{align}
S[a,\phi]\simeq \pm\int^{a}_{0}\mathrm{d}a\cdot v_K \sqrt{
12a^4\left(\frac{3K}{a^2}-V\left( \phi  \right)\right) },
\end{align}
where the kinetic energy of $\phi$ has been neglected, and the two signs correspond to the two independent WKB solutions. More precisely, $\Psi(a, \phi)$ can be a linear combination of similar terms.  Which sign is relevant depends on various proposals. Since the Universe created from nothing has a
finite size $a_\text{ini} \simeq \sqrt{3K/V\left( \phi  \right)}$, we have
the WKB action $S[a_\text{ini},\phi]\simeq\pm 
12 \pi^2 /{V\left( \phi  \right)}$.
Thus, the nucleation probability of the Universe for $\phi$ is given by
two forms~\cite{Vilenkin:1987kf},
\begin{equation}
\mathcal{P}_{0\, \to\, a_\text{ini}}\left( a_\text{ini}, \phi  \right)
\propto \exp\left(\mp \frac{24\pi^2 
}{V\left( \phi  \right)}\right), \label{probability}
\end{equation}
where the minus sign corresponds to the tunneling wave function~\cite{Vilenkin:1984wp},
whereas the plus one to the (Hartle-Hawking) no-boundary wave function~\cite{Hartle:1983ai}.

Let us discuss the implications of the two possibilities of the sign in Eq.~\eqref{probability}. 
For the case of the minus sign, a large value of $V(\phi)$ is probabilistically favored. In this case, it is likely to find $\phi$ on the plateau far away from the minimum, so there is no obstacle to having long-lasting inflation just after the creation of the Universe. Note that for $V_0 \ll 1$, the ratio of the probabilities with different $\phi$ is huge even though the ratio of $V$ is $\mathcal{O}(1)$. Thus, there is no need to rely on the nontrivial cosmological dynamics studied in the preceding sections to realize long-lasting inflation.

More interesting is the case of the plus sign in Eq.~\eqref{probability}, in which a small value of $V(\phi)$ is probabilistically favored.  The conventional view, in this case, is that inflation is quite unlikely to happen.  Note that the formula of the probability is derived assuming the flatness of the potential. Restricting our discussion to the sufficiently flat part of the potential so that the probability formula is valid, the most probable initial state of the Universe is $\phi$ close to the edge of the plateau~\cite{Matsui:2020tyd}.   This implies that the most probable outcome after creation is either (i) the big crunch possibly after bounce(s) and/or a short period of inflation or (ii) a long-lasting inflation after bounce(s).  The conventional view is based on the expectation (i), while the solutions we found in this paper open up the possibility of outcome (ii).  If we consider the conditional probability given long-lasting inflation, the Universe probably experiences single or multiple bounces since otherwise, it would lead to the big crunch with a high probability. This means that the solutions we found in this paper may be highly relevant to the no-boundary proposal. 

The reasoning in the previous paragraph drastically changes when we restore the small positive cosmological constant $\Lambda$. This allows a de Sitter state at the minimum of the potential ($\phi = 0$), and there is no reason to exclude this point when we discuss relative probabilities of the nucleation of the Universe. The probability for the Universe to be created with $\phi = 0$ is exponentially higher than that with $\phi$ close to the edge of the plateau. In this case, it seems to us necessary to use an anthropic argument to explain the reality. Namely, if the universe is created with the observed value of the cosmological constant in the Bunch-Davies vacuum and with the same parameters in the Standard Model of particle physics, there will be no baryons, no galaxies, and no observers. Again, if we consider the conditional probability to have an intellectual observer in the Universe, the nontrivial cosmological dynamics found in this paper will be highly relevant.

\subsection{Comments on perturbations and instabilities}

Similar to the controversy of the probability distribution of the wave function of the Universe at the homogeneous level, the probability distribution of perturbations is also controversial. 
In other words, whether the highest nucleation probability is at the most symmetric configuration so that the homogeneous and isotropic Universe is realized is still under debate in quantum cosmology~\cite{Feldbrugge:2017fcc, DiazDorronsoro:2017hti, Feldbrugge:2017mbc, Feldbrugge:2018gin, DiazDorronsoro:2018wro, Halliwell:2018ejl, Janssen:2019sex, Vilenkin:2018dch, Vilenkin:2018oja, Bojowald:2018gdt, DiTucci:2018fdg, DiTucci:2019dji, DiTucci:2019bui, Lehners:2021jmv, Matsui:2021yte, Martens:2022dtd, Matsui:2022lfj}.  
Recently, it has been claimed that the wave function of small perturbations around the background does not take the Gaussian form and is inconsistent with cosmological observations~\cite{Feldbrugge:2017fcc, Feldbrugge:2017mbc}. In this paper, we simply assume that the perturbations are significantly suppressed after the quantum tunneling of the Universe since, otherwise, the nontrivial cosmological dynamics found in this paper would be impossible. 

There are various instabilities in our solutions.  For example, we have already discussed the tachyonic instabilities of $\phi$ and gauge fields.  There is also the instability of our solutions against a tunneling process of the Universe into nothing.  This is basically the opposite process to the quantum creation of the Universe.\footnote{There are some differences. As we have seen above, various perturbations grow after the birth of the Universe by dissipative processes or by tachyonic instabilities. Even if the Universe is born out of a highly symmetric and homogeneous state in a non-singular way, the final fate of the Universe could be singular due to developed inhomogeneity if it is not smeared by quantum gravity effects. 
} Such nonperturbative decay into nothing was discussed in the context of the classically eternal cyclic cosmology~\cite{Mithani:2011en, Mithani:2014toa, Damour:2019iyi}.
In our scenario, the number of bounces is tightly constrained by the fragmentation due to tachyonic instability particularly when we fit the CMB observations (see Fig.~\ref{fig:quasi-cyclic_dissipation}, in which only one bounce (neglecting small oscillations) occurs), so the relevance of quantum disappearance into nothing seems to be significantly low compared to the eternal cyclic scenarios.

%%%%%%%%%%%%%%%%%%%%%%%%%%%%%%%%%%%%%%%%%%%%%%%%%%%
\section{Conclusions \label{sec:conclusions}}

In this paper, we have studied the classical dynamics of the Universe with an inflaton field $\phi$ whose potential has a flat part and with positive spatial curvature. The flat potential is motivated by the CMB observations while the positive spatial curvature and the initial conditions of our study are motivated by the quantum creation of the Universe from nothing.  In particular, we have studied the cases in which the initial value of $\phi$ is close to the edge of the plateau so that a long-lasting inflation does not directly occur.  Interestingly, there are quasi-cyclic solutions, but the cyclic dynamics are terminated by the tachyonic instability, which is an inevitable, built-in feature of our potential.  Nevertheless, we have found that the quasi-periodic solutions can be turned into viable inflationary solutions by taking into account interactions among $\phi$ and other fields, which are anyway necessary to reheat the Universe after inflation. Despite the nontrivial dynamics, we have not introduced any effects beyond \ac{gr} and we have not violated the \ac{nec} at the classical level. The toy models we have studied include the simplest model with a constant dissipation rate and a more realistic axion-like inflaton model with various thermal effects. In both cases, we have demonstrated that the standard slow-roll cold inflation begins.  As the flat potential is favored by the CMB data, these nontrivial cosmological solutions are consistent with the CMB observations. Further, in the minimal setup, the reheating of the Universe occurs through the interactions that triggered inflation.

The most severe requirement for our mechanism to work is that the inflaton efficiently dissipates before it fragments by the tachyonic instability.  This dissipation should not be too slow since otherwise the inflaton fragments and the big crunch occurs.  At the same time, this dissipation should not be too quick since otherwise, the Universe contracts so rapidly that the kinetic energy of $\phi$ becomes too large for the bounce to take place, again resulting in the big crunch. Therefore, there is an allowed window for the strength of the dissipative interactions of $\phi$.  We have clarified the requirements for the ultraviolet completion of our mechanism with respect to this point.

An interesting aspect of our cosmological scenario is related to the wave function of the Universe and the past of our Universe. \emph{If} the wave function of our Universe is close to the one predicted in the no-boundary proposal rather than that in the tunneling proposals, and if we adopt an inflation model with a flat potential, we can argue that the most likely (in the sense of Bayesian conditional probability) past of our Universe before inflation is the quasi-cyclic period discussed in this paper!  This is because the long-lasting inflation just after the creation of the Universe is exponentially unlikely in the no-boundary proposal and also because the Universe would end with the big crunch without the quasi-cyclic period.

One may expect shift-symmetry breaking terms that restrict the length of the flat part of the scalar potential for various reasons~\cite{Harigaya:2014qza, Abe:2014opa, Buchmuller:2015oma, Broy:2014xwa, Broy:2015qna, Rudelius:2019cfh}. For example, an asymptotically exponential potential is typically predicted in simple setups in string theory whose exponent has presumably a lower bound~\cite{Obied:2018sgi, Ooguri:2018wrx, Denef:2018etk, Garg:2018reu, Andriot:2018wzk, Andriot:2018mav, Bedroya:2019snp, vanBeest:2021lhn}. If the lower bound is sufficiently small in contrast to what Swampland conjectures typically predict~\cite{Obied:2018sgi, Ooguri:2018wrx, Denef:2018etk, Garg:2018reu, Andriot:2018wzk, Andriot:2018mav, Bedroya:2019snp, vanBeest:2021lhn}, or if the small deformation of the potential appears not in the asymptotic region, accelerated expansion, i.e., a bounce as well as inflation is possible.  We have found examples of a parameter set for which a small shift-symmetry breaking terms such as $\delta V_0 \cosh(\phi/\phi_0)$, where $\delta$ is a dimensionless parameter, can slightly extend the parameter space of the emergent inflation after the quasi-cyclic period.  This can be understood as follows.  Without the shift-symmetry breaking terms, there is a maximum of the potential $\max_\phi V(\phi) = V_0$, so once the universe shrinks beyond $\rho(\phi) > V_0$, the kinetic energy of $\phi$ grows substantially, leading to the big crunch.  With the positive shift-symmetry breaking terms, the potential can be unbounded (up to the cutoff scale), so the kinetic energy does not necessarily blow up. More systematic studies on the shift-symmetry breaking terms are left for future work.

Let us comment on the observational aspects of our scenario. The current observations are consistent with the flat Universe~\cite{Ata:2017dya, Planck:2018vyg, Madhavacheril:2023qly}.  In principle, some deviations from the standard $\Lambda$CDM prediction such as the evidence of positive spatial curvature may arise at the largest scale of the closed Universe. This is in principle observable if the total e-folding of inflation is just enough to solve the horizon and flatness problems.  Also, suppression of the CMB power spectrum at low multipoles was discussed in Ref.~\cite{Sloan:2019jyl}. 
In addition, there may be some hope to probe the beginning of the Universe indirectly through the connection between the pre-inflationary physics and the reheating after inflation in our scenario provided that we can observationally test some aspects of the reheating.  This is a unique feature of our scenario.  However, there is some model dependence on how to heat the SM sector.  It is worth studying more about this connection between the pre-inflationary epoch and the reheating epoch.

In this respect, it is interesting to speculate the possibility that the non-Abelian gauge fields discussed in our dissipation channel are identified with some of the gauge fields in the SM.  There are light fermions in the SM, so the phenomenology of pre-inflation may be modified significantly. It will be also interesting if the SM Higgs field can play the role of $\phi$ possibly with the non-minimal coupling to gravity to flatten its canonical potential and some ingredients beyond the SM if necessary.  In these cases, the interactions for reheating the SM sector will have more direct connections with the dissipation mechanisms responsible for the bounces and triggering inflation.

In conclusion, we have found a novel cosmological scenario: the Universe is created from nothing, which is followed by a transient quasi-cyclic period, then inflation occurs due to the dissipation of the inflaton energy, and finally the Universe is reheated by the same dissipative interactions.  We hope to come back to study further implications of this scenario in the future.

%%%%%%%%%%%%%%%%%%%%%%%%%%%%%%%%%%%%%%%%%%%%%%%%%%%
\section*{Acknowledgments}
Our analyses on the \CosmoLattice{} outputs were helped by the Mathematica notebook created by Francisco Torrenti that is available on \href{https://cosmolattice.net/}{the \CosmoLattice{} website}. 
T.T.~thanks Jong-Wan Lee for the discussion on the interpretation of the lattice simulation result and Kunio Kaneta for the discussion on the particle production rate from the inflaton condensate. He thanks the Yukawa Institute for Theoretical Physics (YITP) at Kyoto University; he thanks participants for discussions during the YITP workshop YITP-W-22-11 on ``Progress in Particle Physics 2022''. 
He also thanks participants of the workshop ``100+7 GR \& Beyond: Inflation'' for discussions. 
This work was supported by IBS under the project code, IBS-R018-D1 (A.P.~and T.T.), JSPS Core-to-Core Program (Grant No.~JPJSCCA20200002) (F.T.),  JSPS KAKENHI Grants No.~22KJ1782 (H.M.), No.~23K13100 (H.M.),
No.~20H01894 (F.T.), and No.~20H05851 (F.T.). This article is based upon work from COST Action COSMIC WISPers CA21106,  supported by COST (European Cooperation in Science and Technology).

%%%%%%%%%%%%%%%%%%%%%%%%%%%%%%%%%%%%%%%%%%%%%%%%%%%
\appendix
%%%%%%%%%%%%%%%%%%%%%%%%%%%%%%%%%%%%%%%%%%%%%%%%%%%
\section{Analysis of cyclic solutions with oscillation average\label{sec:osc_avg}}

\begin{figure}[btph] 
\begin{center}
\includegraphics[width=0.7 \columnwidth]{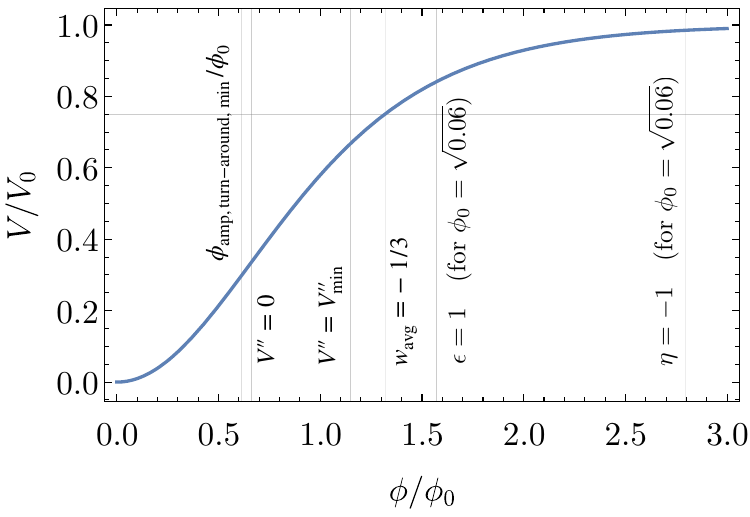}
\caption{The scalar potential (blue line) and various characteristic field values (thin gray vertical lines). From left to right, the vertical lines denote the following field values: (1) the minimum amplitude of $\phi$ when the Universe turns around, (2) the point where the curvature of the potential vanishes, (3) the point where the tachyonic instability is strongest, (4) the average equation-of-state parameter has the critical value $-1/3$ dividing bounce and turn-around, (5) the  point where the first slow-roll parameter $\epsilon = 1$, and (6) the point where the second slow-roll parameter $\eta= -1$. For (5) and (6), $\phi_0 = \sqrt{0.06}$ is chosen as a fiducial value.}
\label{fig:V}
\end{center}
\end{figure}

In this Appendix, we summarize the basic properties of our example potential~\eqref{potential}, $V = V_0 \tanh^2 (\phi / \phi_0)$, and analytic results for the dynamics of $\phi(t)$ and $a(t)$ under some approximations.  The shape of the potential is shown in Fig.~\ref{fig:V} with some characteristic field values introduced in the following. 

The effective mass of the scalar field $\phi$ depends on its field value.  At the minimum, it is given by $m^2 = 2 V_0 / \phi_0^2$.  The curvature of the potential vanishes at
$|\phi_{m^2=0}| /\phi_0 = \arcosh(2)/2 \approx 0.658479$.
At this point, the height of the potential is
$V(\phi_{m^2 = 0}) = V_0/3$.  The tachyonic mass becomes strongest when $V''' = 0$. This occurs at $|\phi_{m^2_\text{min}}|/\phi_0 = \log(5 + 2 \sqrt{6})/2 \approx 1.14622$.
The strongest tachyonic mass is $V''(\phi_{m^2_\text{min}}) = -  m^2 /3$. At this point, the height of the potential is $V(\phi_{m^2_\text{min}}) = 2 V_0 / 3$. These as well as other field values are summarized in Table~\ref{tab:field_values}.

%%%%%%
\begin{table}[ptbh]
\begin{center}
\caption{Characteristic field values and the values of the potential (in the oscillation-average approximation). The values in parentheses are oscillation-averaged quantities. See the main text for analytic expressions. For comparison, field values characterized by the slow-roll parameters $\epsilon \equiv (V'/V)^2/2$ and $\eta \equiv V''/V$ are also shown.}
    \begin{tabular}{|l|c|c|} 
    \hline
      Description  &  $|\phi|/\phi_0$ \, ($\phi_\text{amp}/\phi_0$) & $V(\phi)/V_0$ \, ($\rho / V_0$) \\ \hline \hline
      Lower bound for turn-around & $(0.609378)$ & ($0.295598$) \\ \hline 
      Vanishing effective mass & $0.658479$ & $1/3$ \\ \hline
      Strongest tachyonic instability & $1.14622$ & $2/3$ \\ \hline 
      Critical value for bounce vs turn-around & ($1.31696$) & $(3/4)$ \\ \hline
      $\epsilon = 1$ for $\phi_0 = \sqrt{0.06}$ & $1.56979$ & 0.840874 \\ \hline
    $\eta = -1$ for $\phi_0 = \sqrt{0.06}$ & $2.79300$ & 0.985112 \\    
      \hline
    \end{tabular}
    \label{tab:field_values}
\end{center}
\end{table}
%%%%%%
%%%%%%%%%%%%%%%%%%
%%%%%%%%%%%%%%%%%%

In the flat spacetime limit, the equation of motion can be analytically solved~\cite{Tomberg:2021bll}:
\begin{align}
    \phi(t) = \phi_0 \arsinh \left( \frac{\cos \left( \pi (t - t_0) / \Thalf \right)}{\sqrt{V_0/\rho - 1}} \right),
\end{align}
where $t_0$ is a parameter governing the origin of time $t$,  $\Thalf = \frac{\pi \phi_0}{\sqrt{2 (V_0 - \rho)}}$ is the half-period of the oscillations of $\phi$, and $\rho$ is the energy density, which is a constant of motion in the flat limit. Letting $\phi_\text{amp}$ denote the amplitude of the oscillations, $\rho= V_0 \tanh^2(\phi_\text{amp}/\phi_0)$. 

The equation-of-state parameter $w = P / \rho$ is an important quantity since its value at the moment of $H = 0$ determines if the Universe expands or contracts at the next moment as dictated by $\ddot{a}/a = - \rho(1 + 3 w) / 6$.  
It crosses the critical value $-1/3$ at
\begin{align}
    \frac{|\phi_{w = -1/3}|}{\phi_0} =&  \artanh \left(\sqrt{\frac{2}{3}} \tanh \left( \frac{\phi_\text{amp}}{\phi_0} \right) \right).
\end{align}
$|\phi_{w = -1/3}|$ is upper bounded by $|\phi_{m^2_\text{min}}|$.  When the amplitude is large $\phi_\text{amp} \gtrsim \phi_0$, it saturates the bound $|\phi_{w=-1/3} |\approx |\phi_{m^2_\text{min}}|$.  When the amplitude is small $\phi_\text{amp} \lesssim \phi_0$, the bounce can occur only around the edges of the oscillations, $|\phi_{w=-1/3}| \approx \phi_\text{amp}$.

In the following, we assume that the scalar field oscillates around the minimum and that it is a good approximation to take the oscillation average. This is a better approximation for a smaller $\phi_0$ and a smaller $\phi_\text{amp}/\phi_0$.  We do coarse-graining and consider the effective fluid having the equation-of-state parameter $w_\text{avg} = \langle w \rangle_\text{osc}$ where the subscript osc means the oscillation average. Thereafter, we consider the relatively slow dynamics of the scale factor.  Note that, in reality, when the field amplitude becomes sufficiently large for a given $\phi_0$, the curvature dilutes sufficiently, $H$ becomes non-negligible, and inflation occurs.  Once (long-lasting) inflation occurs, the assumption of the oscillation average is completely broken.   

Using the technique of Refs.~\cite{Karam:2021sno, Tomberg:2021bll}, $w_\text{avg}$ can be written as
\begin{align}
   w_\text{avg}= \frac{-2 + 2 \sqrt{1-\rho/V_0}+\rho/V_0}{\rho/V_0} . \label{w_avg}
\end{align}
We can invert this to obtain
\begin{align}
    \frac{\rho}{V_0} = \frac{- 4 w_\text{avg}}{(1 - w_\text{avg})^2}.
\end{align}

%%%%%%%%%%%%%%%%%%
\begin{figure}[htbp] 
\begin{center}
\includegraphics[width=0.6 \columnwidth]{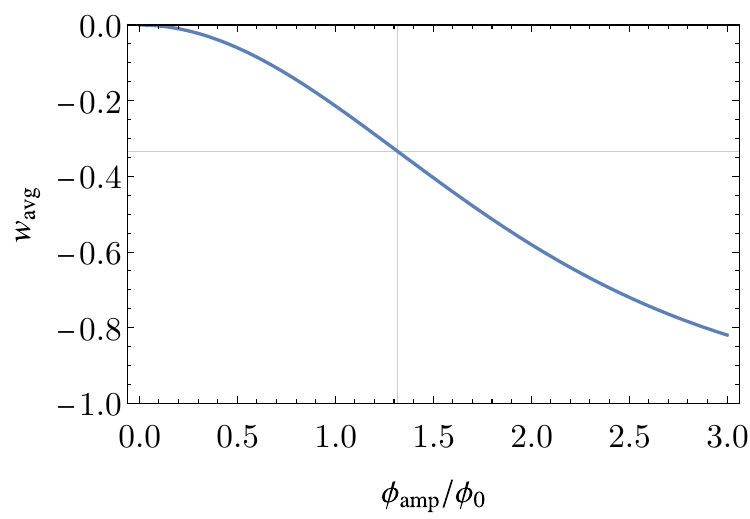}
\caption{The coarse-grained equation-of-state parameter $w_\text{avg} = \langle w \rangle_\text{osc}$ after the oscillation average. The thin horizontal line denotes the critical value $w_\text{avg} = -1/3$, and the thin vertical line denotes the corresponding amplitude $\phi_{\text{amp},w_\text{avg}=-1/3}$.}
\label{fig:w_avg}
\end{center}
\end{figure}
%%%%%%%%%%%%%%%%%%

$\phi_\text{amp}$ and $w_\text{avg}$ ($-1 < w_\text{avg} \leq 0$) are related as follows:
\begin{align}
    \frac{|\phi_\text{amp}|}{\phi_0} =  \log \frac{(1 + \sqrt{-w_\text{avg}})^2}{1 + w_\text{avg}}. 
\end{align}
In particular, the value of $w_\text{avg}$ becomes $-1/3$ when 
\begin{align}
    \frac{|\phi_{\text{amp}, w_\text{avg}=-1/3}|}{\phi_0} =  \log ( 2 + \sqrt{3}) \approx  1.31696.
\end{align}
At this point, $\rho/V_0 = 3/4$.
The dependence $w_\text{avg}(\rho(\phi_\text{amp}))$ is shown in Fig.~\ref{fig:w_avg}. 

\begin{figure}[htbp] 
\begin{center}
\includegraphics[width=0.49 \columnwidth]{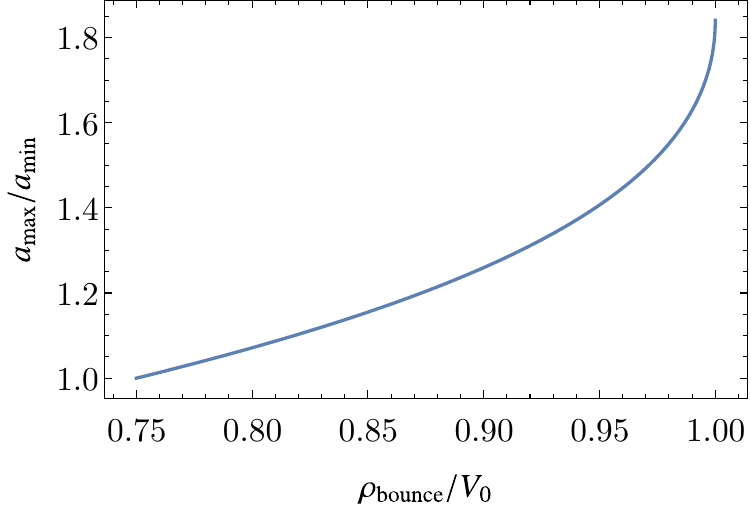}
\includegraphics[width=0.49 \columnwidth]{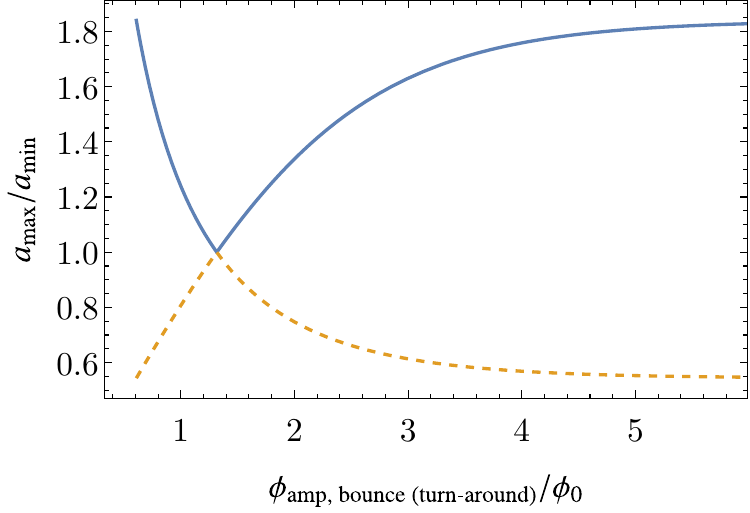}
\caption{ The ratio $a_\text{max}/a_\text{min}$ of the minimum and maximum scale factors with the oscillation average approximation. The maximum ratio is about 1.83929. Left:\, in terms of $\rho_\text{avg, bounce}$. Right:\, in terms of $\phi_\text{amp, bounce}$. The dashed line denotes the corresponding $a_\text{min}/ a_\text{max}$. The value of $\phi_\text{amp}/\phi_0$ corresponds to that at the bounce (turn-around) time when it is larger (smaller) than the position of the kink at around 1.31696, respectively. }
\label{fig:a_max_ratio}
\end{center}
\end{figure}

Combining Eq.~\eqref{w_avg} with $\dot{\rho}+3(1+w_\text{avg}(\rho)) H \rho = 0$, we can relate $\rho$ and $a$.  Integrating it, we have
\begin{align}
    \left( \frac{a}{a_*} \right)^{-3} = \frac{1 - \sqrt{1 - \frac{\rho_\text{avg}}{V_0}}}{1 - \sqrt{1 - \frac{\rho_\text{avg, *}}{V_0}}}, \label{relation_a_rho}
\end{align}
where the star denotes a reference scale.
In particular, from the condition $\rho_\text{bounce} a_\text{min}^2 = \rho_\text{turn-around} a_\text{max}^2 (= 3 K)$, 
\begin{align}
    \left( \frac{a_\text{min}}{a_\text{max}} \right)^3 = \frac{1 - \sqrt{1 - \frac{\rho_\text{avg, bounce}}{V_0}\left( \frac{a_\text{min}}{a_\text{max}} \right)^2}}{1 - \sqrt{1 - \frac{\rho_\text{avg, bounce}}{V_0}}},
\end{align}
where the subscripts bounce and turn-around represent that the quantities are evaluated at the bounce and turn-around, respectively, and $a_\text{min}$ and $a_\text{max}$ are the minimum and maximal values of the scale factor, respectively. 
This can be numerically solved for given $\rho_\text{avg, bounce}/V_0$ or $\rho_\text{avg, turn-around}/V_0$. It is easier to solve it for $\rho_\text{avg, bounce}$ in terms of $a_\text{max}/a_\text{min}$:
\begin{align}
    \frac{\rho_\text{avg, bounce}}{V_0} =\left. \frac{4r^3 (1 + r + r^2)}{(1+r + r^2 + r^3)^2} \right |_{r \equiv \frac{a_\text{max}}{a_\text{min}}}.
\end{align}
This relation is plotted in the left panel in Fig.~\ref{fig:a_max_ratio}. Using the relation $\rho= V_0 \tanh^2(\phi_\text{amp}/\phi_0)$, it can also be plotted in terms of $\phi_\text{amp, bounce}$ [see the right panel in Fig.~\ref{fig:a_max_ratio}].
In the formal limit of the infinite amplitude (which is of course impossible in the oscillation average approximation as it triggers inflation), 
\begin{align}
    \lim_{\rho_\text{avg, bounce}\to V_0} \frac{a_\text{max}}{a_\text{min}}  = & \frac{1 + \sqrt[3]{19 - 3 \sqrt{33}} + \sqrt[3]{19+3\sqrt{33}} }{3} \nonumber \\
        \approx & 1.83929.
\end{align}  In this case, the minimum value of $\phi_\text{amp}$ at the turn-around time is realized. In other words, if the initial value of $\phi/\phi_0$ is smaller than a critical value, there is no cyclic solution in the regime in which the oscillation average is valid.  The critical value is
\begin{align}
    \min \frac{\phi_\text{amp, turn-around} }{\phi_0} =& \artanh \left( \frac{3}{1 + \sqrt[3]{19 - 3 \sqrt{33}} + \sqrt[3]{19+3\sqrt{33}} } \right) \nonumber \\
    \approx & 0.609378.
\end{align}
At this point, $\rho/V_0 \approx 0.295598$. 

%%%%%%%%%%%%%%%%%%
%%%%%%%%%%%%%%%%%%

We have obtained the relation~\eqref{relation_a_rho} between $a$ and $\rho_\text{avg}$.  To relate it to time $t$, we need to substitute $\rho=\rho(a)$, i.e.,
\begin{align}
    \frac{\rho_\text{avg}}{V_0} 
    =& 2 \left( 1 - \sqrt{1 - \frac{\rho_\text{avg, *}}{V_0}} \right) \left(\frac{a_*}{a} \right)^3 - \left( 1 - \sqrt{1 - \frac{\rho_\text{avg, *}}{V_0}} \right)^2 \left(\frac{a_*}{a} \right)^6, \label{rho_avg(a)}
\end{align}
into the Friedmann equation, $H^2 = \frac{\rho}{3} - \frac{K}{a^2}$ and integrate it.
It is convenient to take the reference time * as the bounce time (or, alternatively, the turn-around time) so that $K$ can be solved. The differential equation we should solve is
\begin{align}
    \left( \frac{\dot{a}}{a} \right)^2 = \frac{2 V_0 X_*}{3} \left( \left(\frac{a_*}{a} \right)^3 - \frac{X_*}{2} \left(\frac{a_*}{a} \right)^6 - \left( 1- \frac{X_*}{2}\right) \left(\frac{a_*}{a} \right)^2 \right),
\end{align}
or 
\begin{align}
    \frac{\ddot{a}}{a} =& - \frac{V_0}{6} \left( 4 \left( 2 X_* \left( \frac{a_*}{a} \right)^{3} - \frac{X_*}{2}\left( \frac{a_*}{a} \right)^{6} \right) -6 + 6 \sqrt{1 - \left( 2 X_* \left( \frac{a_*}{a} \right)^{3} - \frac{X_*}{2}\left( \frac{a_*}{a} \right)^{6} \right)}  \right), \label{Friedmann_eq_avg2}
\end{align}
where $X_* \equiv 1 - \sqrt{1 - \rho_\text{avg, *}/V_0}$ for compact notation.

%%%%%%%%%%%%%%%%%%
\begin{figure}[htbp] 
\begin{center}
\includegraphics[width=0.6 \columnwidth]{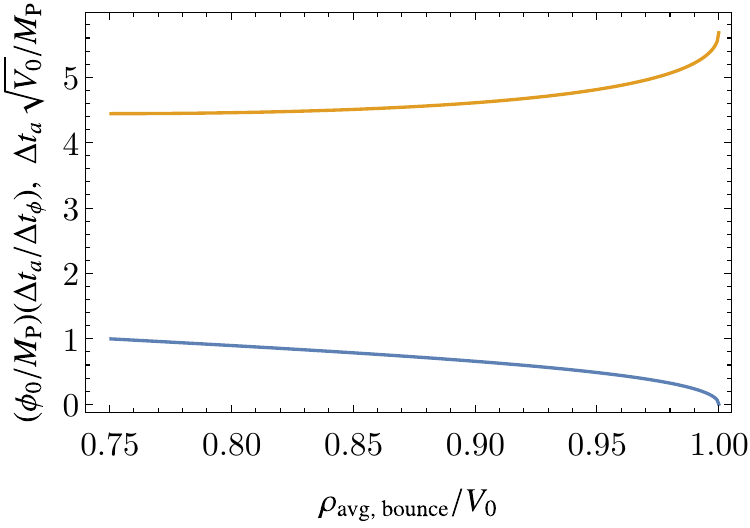}
\caption{The half period $\Thalfa$ of the single oscillation of the scale factor $a$ in the cyclic solution with scalar-field oscillation average approximation.  The top (orange) line shows $\Thalfa$ in units of $(\sqrt{V_0}/M_\text{P})^{-1}$.  The bottom (blue) line shows $\Thalfa / \Thalf$, i.e., how many scalar-filed oscillations occur in a single scale-factor oscillation in units of $(M_\text{P}/\phi_0)$. }
\label{fig:period}
\end{center}
\end{figure}
%%%%%%%%%%%%%%%%%%

The half period $\Thalfa$ of the oscillation of $a$ in the cyclic solution in units of $M_\text{P}/\sqrt{V_0}$ is obtained numerically, and it is shown as the top (orange) solid line in Fig.~\ref{fig:period}.  Its value is around 5.  The number of oscillations of the scalar field within one cycle of the scale-factor oscillation is given by $\Thalfa/\Thalf$.  
 The ratio $\Thalfa/ \Thalf$ is proportional to $1/\phi_0$, so $ \phi_0 \Thalfa/\Thalf$ is shown as the bottom (blue) solid line in Fig.~\ref{fig:period}.  Its value is around 1.

\begin{figure}[htbp] 
\begin{center}
\includegraphics[width=0.6 \columnwidth]{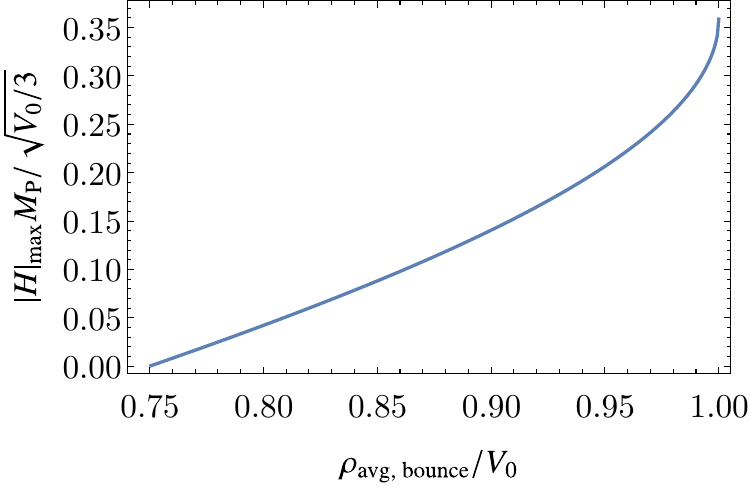}
\caption{The oscillation amplitude of the Hubble parameter $|H|_\text{max}$ as a function of the averaged energy density $\rho_\text{avg}$ evaluated at the bounce point.  This is a measure of how much cancellation happens between $\rho$ and $\rho_\text{curv}$.  The vertical axis is normalized in such a way that it equals 1 if $|\phi|$ goes to infinity and inflation starts (so that $\rho_\text{curv}$ becomes negligible).}
\label{fig:|H|_max}
\end{center}
\end{figure}

In the cyclic regime, the Hubble parameter oscillates around 0.  It is useful to know its typical value, i.e., the amplitude of the oscillation. The maximum of the absolute value of the Hubble parameter $|H|_\text{max}$ is shown in Fig.~\ref{fig:|H|_max}. Because of the cancellation between the positive energy density and the negative curvature term, the typical amplitude is suppressed compared to the inflationary Hubble scale. 

%%%%%%%%%%%%%%%%%%
%%%%%%%%%%%%%%%%%%

%%%%%%%%%%%%%%%%%%%%%%%%%%%%%%%%%%%%%%%%%%%%%%%%%%%%%%%
\section{Details of lattice calculations \label{sec:lattice}}
We use \CosmoLattice~\cite{Figueroa:2020rrl, Figueroa:2021yhd} for the classical lattice simulation to assess the significance of the tachyonic instability of $\phi$. 

\paragraph{Lattice topology}
Although we are interested in the closed Universe, it is beyond the scope of our work to modify the spatial lattice structure from the torus $\mathbb{T}^3$ (because of the periodic boundary condition) to the sphere $\mathbb{S}^3$ as mentioned in the main text.  Instead, we discuss the infrared cutoff scale due to the closed topology below. 

\paragraph{Time variable} 
In \CosmoLattice, the ``$\alpha$-time'' $\mathrm{d}\eta := a^{-\alpha_{\mathcal{C}\mathcal{L}}} \mathrm{d} t$ is introduced.\footnote{
To distinguish it from the gauge coupling $\alpha$, we attach the subscript $\mathcal{C}\mathcal{L}$.
} For example, $\alphaCL = 0$ and $1$ correspond to the usual cosmic time and the conformal time.  
 In our work, we set $\alphaCL = 0$ so that the $\alpha$-time is nothing but the usual time $\eta = t$. Below, we keep this parameter since it may be useful for other applications. 
 
\paragraph{Dimensionless variables}
In \CosmoLattice, dimensionful quantities are made dimensionless by the introduction of reference scales $f_*$, which has the dimension of a canonically normalized field, and $\omega_*$, which has the dimension of frequency. A tilde on top of a variable denotes the dimensionless quantity rescaled by $f_*$ and/or $\omega_*$.  
 We choose 
 \begin{align}
    f_* =& \phi_0 & \text{and}& & 
    \omega_* = \frac{\sqrt{2 V_0}}{\phi_0}.
 \end{align}
With this choice, in particular, $\tilde{V} = V/(f_*^2 \omega_*^2) = \frac{1}{2} \tanh^2 \tilde{\phi}$ becomes simple, and $\tilde{t}  = \omega_* t = (\sqrt{2V_0}/\phi_0) t$ is used for the horizontal axes of Figs.~\ref{fig:cyclic}, \ref{fig:lattice}, \ref{fig:quasi-cyclic_dissipation}, and \ref{fig:thermal_quasi-cyclic_dissipation}.

\paragraph{Addition of the spatial curvature term}
We need to modify the codes of the \CosmoLattice{} to include the curvature term and run it in the self-consistent expansion mode. The Friedmann equations in terms of the program variables are
\begin{align}
    \left(\frac{a'}{a} \right)^2 = & \frac{a^{2\alphaCL}}{3}\left( \frac{f_*}{M_\text{P}} \right)^2 \left( \langle \tilde{\rho}_\text{kin} + \tilde{\rho}_\text{grad} + \tilde{V}\rangle  + \tilde{\rho}_K \right) , \label{Friedmann_dimensionless} \\
    \frac{a''}{a} =& \frac{a^{2\alphaCL}}{3} \left( \frac{f_*}{M_\text{P}} \right)^2 \left(  (\alphaCL - 2) \langle \tilde{\rho}_\text{kin} \rangle + \alphaCL \langle \tilde{\rho}_\text{grad} \rangle + (\alphaCL + 1) \langle \tilde{V} \rangle + \alphaCL \tilde{\rho}_K  \right),
\end{align}
where the prime is the derivative with respect to $\tilde{\eta} = \omega_* \eta$, a bracket denotes the volume average, $\tilde{\rho}_\text{kin}$ and $\tilde{\rho}_\text{grad}$ are the kinetic energy density and the gradient energy density normalized as above. These and the curvature energy density are normalized by the multiplication with $f_*^{-2}\omega_*^{-2}$. 

\CosmoLattice{} solves the second Friedmann equation (the equation for $a''$), and the first Friedmann equation (that for $a'$) is used to monitor the constraint equation (``energy conservation'').  When we choose $\alphaCL = 0$, the second Friedmann equation is not modified by the spatial curvature.  Instead, we modify the initial condition, which is derived by using the first Friedmann equation.

\paragraph{Infrared cutoff scales}
The minimum wavenumber in \CosmoLattice{} is defined as 
\begin{align}
    k_\text{IR, lattice} = \frac{2\pi}{L} = \frac{2 \pi}{N_{\mathcal{C}\mathcal{L}} \, \delta x},
\end{align}
where $L$ is the length of the simulation box in one dimension, $N_{\mathcal{C}\mathcal{L}}$ is the number of lattice sites in one dimension, and $\delta x$ is the lattice spacing.  The subscripts ``lattice'' are added to distinguish it from similar ones introduced below.   Practically, $N_{\mathcal{C}\mathcal{L}}$ and $\tilde{k}_\text{IR, lattice} = k_\text{IR, lattice} / \omega_*$ are input parameters, while $L$ and $\delta x$ are implicit. $k_\text{IR, lattice}$ should be smaller than physically interesting scales. 

One such candidate is the peak scale of the tachyonic instability $k_\text{peak}$.  This depends on $\phi_\text{amp}$~\cite{Tomberg:2021bll}. In the asymptotic limit $\phi_\text{amp} \to \infty$ and neglecting the cosmic expansion/contraction, $k_\text{peak}/a \approx 3.54 / \Thalf$~\cite{Tomberg:2021bll}.  For generic $\phi_\text{amp}$, we may approximately use 
\begin{align}
    \frac{k_\text{peak}}{a(t)} \approx \frac{3}{\Thalf}. 
\end{align}

Another important scale is the size of the closed Universe itself.  Let us assume the simplest closed topology, i.e., three-dimensional sphere $\mathbb{S}^3$.  
 The size $ a/\sqrt{K}$ of the Universe at the transition from the Euclidean regime into the Lorentzian regime is 
\begin{align}
   \frac{a}{\sqrt{K}} \simeq \sqrt{\frac{3 M_\text{P}^2}{\rho_\text{avg}}}.
\end{align}
The fundamental IR cutoff in our setup should be 
\begin{align}
    \frac{k_{\text{IR}, \mathbb{S}^3}}{a} = \frac{2\pi \sqrt{K}}{\pi a} \simeq 2 \sqrt{\frac{\rho_\text{avg}}{3 M_\text{P}^2}},
\end{align}
where $a$ is evaluated at the nucleation of the Universe. \CosmoLattice{} assumes $a=1$ at the initial time.

We should choose $k_\text{IR, lattice}$ to satisfy $k_{\text{IR}, \mathbb{S}^3} \leq k_\text{IR, lattice}$ since otherwise we compute physically non-existent modes.  Also, we should choose $k_\text{IR, lattice}$ to satisfy $k_\text{IR, lattice} \leq k_\text{peak}$ since otherwise we miss the most important tachyonic mode. Therefore, when $k_{\text{IR}, \mathbb{S}^3} \leq k_\text{peak}$, it is required that $k_{\text{IR}, \mathbb{S}^3} \leq k_\text{IR, lattice}\leq k_\text{peak}$.

\paragraph{Evolver and input parameters}
In the calculations for Fig.~\ref{fig:lattice}, we have adopted the default evolver (the velocity Verlet algorithm: \texttt{VV2}) and we have not turned on the optional ultraviolet cutoff parameter. Input parameters include $N_{\mathcal{C}\mathcal{L}} = 32$, $\tilde{k}_{\text{IR, lattice}} = 0.1$, and the time step $\mathnormal{\Delta} \tilde{t}= 0.01$.  Note that this choice of $\tilde{k}_\text{IR, lattice}$ slightly violates $k_{\text{IR}, \mathbb{S}^3} \leq k_\text{IR, lattice}$, but the lower $k$ modes than the tachyonic peak mode do not significantly affect the dynamics.  The choice is motivated to have the peak mode $k_\text{peak}$ around the center of our window $[k_\text{IR, lattice} , N_{\mathcal{C}\mathcal{L}} \, k_\text{IR, lattice}]$ in the logarithmic scale for a conservative estimate of the time scale that is not affected by the tachyonic instability. 

We have computed it also with the 10 times finer time step.  In this case, the violation of energy conservation becomes significantly improved but the solutions themselves do not change much. Also, we have studied with the $2^3 = 8$ finer lattice ($N_{\mathcal{C}\mathcal{L}}=64$) and again observed only minor changes. 

The calculation breaks down at $\tilde{t} = t \omega_* \approx 77.1$ when the kinetic energy of $\phi$ becomes large.

\paragraph{Initialization}
\CosmoLattice{} first initializes $a$ and $a'$, then initializes other fields and their perturbations, and then corrects $a'$ by using the first Friedmann equation. Because of this, even when the maximum possible value of the curvature, which should ideally lead to $a'=0$, is input, $a'$ becomes around $\mathcal{O}(10^{-8})$ or $\mathcal{O}(10^{-7})$ depending on the parameters such as $V_0$.

\begin{figure}[htbp] 
\begin{center}
\includegraphics[width=0.49 \columnwidth]{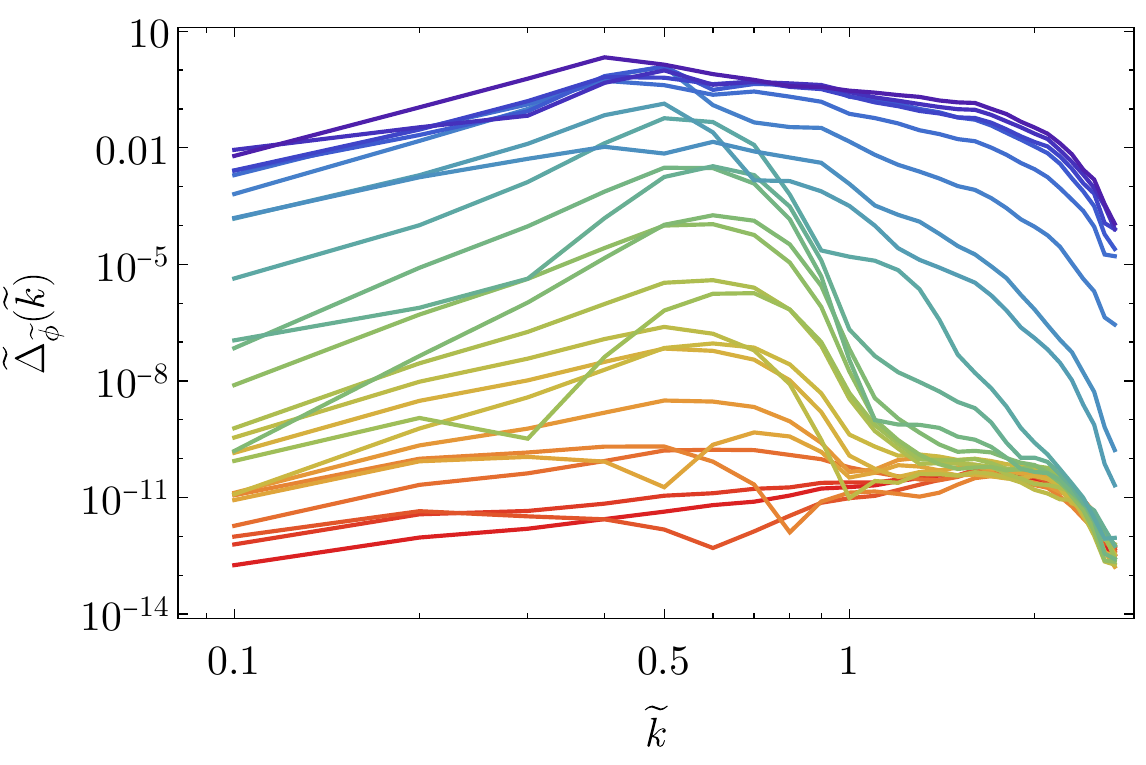}
\includegraphics[width=0.49 \columnwidth]{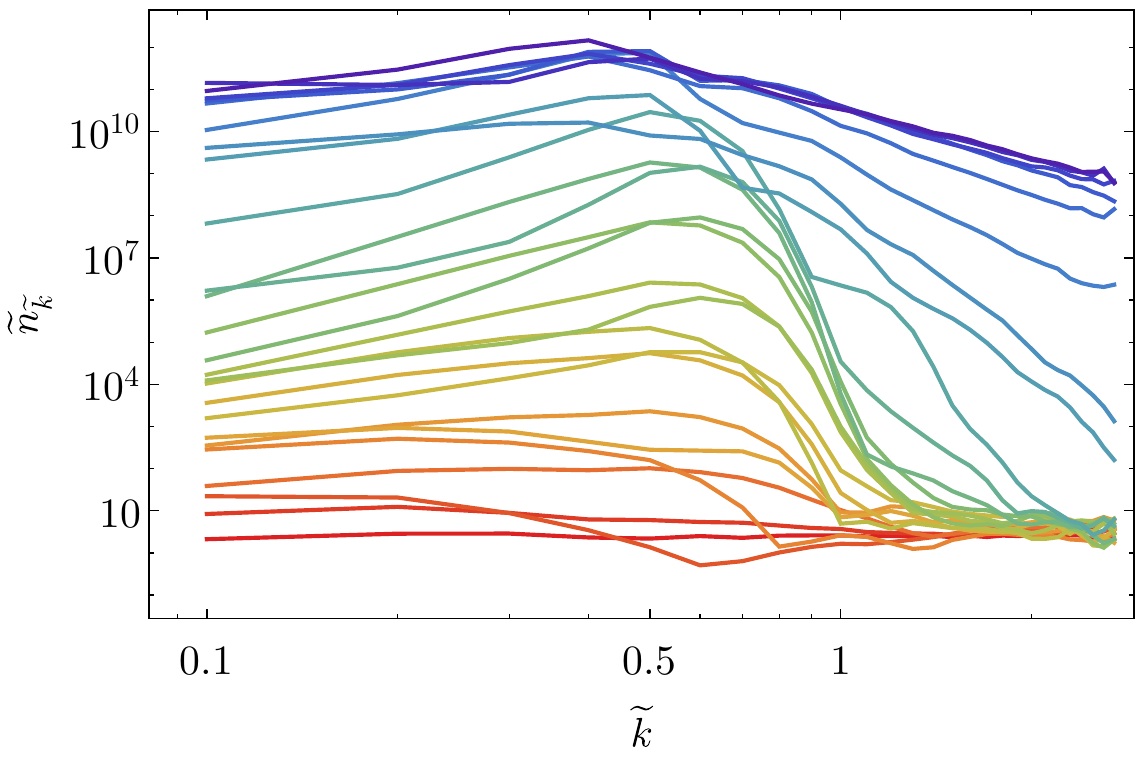}
\caption{ The spectra of perturbations. Time evolves from the bottom (red) line (just after creation) to the top (purple) line (just before the big crunch). Left:\, the dimensionless power spectrum $\widetilde{\Delta}_{\widetilde{\phi}}(\widetilde{k})$ of $\widetilde{\phi}=\phi/f_*$.  Right:\, the spectrum of the dimensionless phase space number density $\widetilde{n}_{\widetilde{k}}$. For more details of the definitions of these lattice quantities, see the manual of the \CosmoLattice{}~\cite{Figueroa:2021yhd}.}
\label{fig:spectra}
\end{center}
\end{figure}

\paragraph{Spectra of perturbations}
The time evolution of the spectra of perturbations is shown in Fig.~\ref{fig:spectra}. The tachyonic peak structure is well resolved. 

%%%%%%%%%%%%%%%%%%
%%%%%%%%%%%%%%%%%%

%%%%%%%%%%%%%%%%%%%%%%%%%%%%%%%%%%%%%%%%%%%%%%%%%%%
\small
\bibliographystyle{utphys}
\bibliography{main_rev2.bib}

\end{document}